\documentclass[preprint]{aastex}

\newcommand\cm{{\rm cm}}
\newcommand\G{{\rm G}}
\newcommand\K{{\rm K}}

\newcommand\Myr{{\rm Myr}}

\newcommand\kms{{\rm km\, s^{-1}}}
\newcommand\pc{{\rm pc}}

\newcommand\simgt{\lower.5ex\hbox{$\; \buildrel > \over \sim \;$}}
\newcommand\simlt{\lower.5ex\hbox{$\; \buildrel < \over \sim \;$}}

\shortauthors{Ostriker et al}
\shorttitle{Structure of Model Clouds}

\begin{document}

\title{Density, Velocity, and Magnetic Field Structure in \\
Turbulent Molecular Cloud Models}

\author{Eve C. Ostriker$^{1,2}$, James M. Stone$^1$, and Charles F. Gammie$^{3}$}
\affil{$^1$Department of Astronomy, University of Maryland \\
College Park, MD 20742-2421}
\affil{$^2$Institute for Theoretical Physics, UCSB, Santa Barbara, CA 93106}
\affil{$^3$Center for Theoretical Astrophysics, University of Illinois\\
1002 W.Green St. Urbana,IL 61801}
\email{ostriker@astro.umd.edu, jstone@astro.umd.edu, gammie@uiuc.edu}

\begin{abstract}

We use three-dimensional (3D) numerical magnetohydrodynamic
simulations to follow the evolution of cold, turbulent, gaseous
systems with parameters chosen to represent conditions in giant
molecular clouds (GMCs).  We present results of three model cloud
simulations in which the mean magnetic field strength is varied
($B_0=1.4-14\ \mu\G$ for GMC parameters), 
but an identical initial turbulent velocity
field is introduced. 
We describe the energy evolution, showing that 
(i) turbulence decays rapidly, with the turbulent energy reduced by a
factor two after 0.4-0.8 flow crossing times ($\sim 2-4\ \Myr$ for
GMC parameters), and (ii) the magnetically supercritical cloud
models gravitationally collapse after time $\approx 6\ \Myr$, while
the magnetically subcritical cloud does not collapse.

We compare density, velocity, and magnetic field structure in
three sets of model ``snapshots'' with
matched values of the Mach number ${\cal M}
\approx 9,7,5$.  We show that 
the distributions of volume density and column density are both
approximately log-normal, with mean mass-weighted volume
density a factor $3-6$ times the unperturbed value, but mean 
mass-weighted column density only a factor $1.1-1.4$ times the 
unperturbed value.  
We introduce a spatial binning algorithm to investigate the dependence
of kinetic quantities on spatial scale for regions of column density 
contrast (ROCs) 
on the plane of the sky. We show that the average velocity dispersion
for the distribution of ROCs is only weakly correlated with scale,
similar to mean size-linewidth distributions for clumps
within GMCs.  We find that ROCs are often
superpositions of spatially unconnected regions that cannot easily be
separated using velocity information; we argue that the 
same difficulty may affect observed GMC clumps.  
We suggest that it may be possible to deduce the mean 3D
size-linewidth relation using the lower envelope of the 2D
size-linewidth distribution.  
We analyze magnetic field structure, and 
show that in the high density regime
$n_{H_2}\simgt 10^3 \cm^{-3}$, total magnetic field strengths increase
with density with logarithmic slope $\sim 1/3 -2/3$. 
We find that mean
line-of-sight magnetic field strengths may vary widely across a
projected cloud, and are not positively correlated with column
density.  We compute simulated interstellar polarization maps 
at varying observer orientations, and determine that the 
Chandrasekhar-Fermi 
formula multiplied by a factor $\sim 0.5$ yields a
good estimate of the plane-of sky magnetic field strength, provided
the dispersion in polarization angles is $\simlt 25^\circ$.

\end{abstract}


\section{Introduction}

Since the identification of cold interstellar clouds 
in radio molecular lines, observational campaigns in many wavelengths 
have provided an increasingly detailed and
sophisticated characterization of their structural properties.  These
clouds are self-gravitating entities permeated by magnetic fields and
strongly supersonic turbulence; the observational properties of giant
molecular clouds (GMCs) are summarized, for example, by
\cite{bli93,wil00,eva99}.  Although it has long been appreciated by
theorists that turbulence and magnetic fields must play a decisive role
in cloud dynamics (e.g.  \cite{mes56, shu87, mck93, shu99, mck99}), much of the
theoretical emphasis has been on evolutionary models in which the
effects of turbulent magnetohydrodynamics (MHD) is modeled rather than
treated in an explicit fashion.  

Recent advances in computer hardware and development of robust
computational MHD algorithms have now made it possible to evolve
simplified representations of molecular clouds using direct numerical
simulations.  Fully nonlinear, time-dependent, MHD integrations can test
theoretical ideas about the roles of turbulence and magnetic fields in
cloud evolution, and also make it possible to investigate how turbulence
affects the structural properties of clouds.  Progress in the
rapidly-developing field of simulations of GMC turbulence is reviewed
by, e.g., \cite{vaz00}.

This is the fourth in a series of papers
\citep{gam96,sto98,ost99}[Papers I-III] investigating the dynamics of
turbulent, magnetized, cold clouds using direct numerical simulations.
The previous papers presented several results on energetics and
overall cloud evolution.  They showed that (1) MHD turbulence can
delay gravitational 
collapse along the mean magnetic field in one-dimensional models
since dissipation is slow (Paper I); however (2) in higher dimensional
models dissipation occurs on the flow crossing timescale $t_f$ (Paper
II); as a consequence, (3) the fate of a cloud depends on whether its
mass-to-magnetic flux ratio is subcritical or supercritical,
independent of the initial turbulent excitation, provided that
turbulence is not steadily driven (Paper III).

Some important astrophysical implications of these results are
that (1) star formation in turbulent clouds may be initiated rapidly,
essentially on a flow crossing timescale; and 
(2) models that rely on slowly dissipating turbulence to
support GMCs against collapse do not appear to be viable.  
One is still faced with the problem of avoiding the excessive Galactic 
star formation rate that would result from the
collapse and fragmentation of the whole cold component of the 
ISM within its gravitational 
free-fall time (comparable to its flow crossing time; \cite{zuc74}).  
This requires
either (a) limitation of the star formation rate in individual
clouds (if self-gravitating clouds are long-lived after formation), or
(b) limitation of the lifetimes of self-gravitating clouds. Both of these
effects may be important. Processes
that contribute to limiting star formation {\it rates} in individual clouds 
include turbulent feedback from star formation, transmission of
turbulence from the larger-scale ISM, or 
a large (subcritical) mean magnetization of clouds.  The first two 
of these processes, together with destabilizing 
environmental factors such as enhanced 
galactic shear outside spiral arms, contribute to limiting 
{\it lifetimes} of individual clouds.

Other workers have independently used simulations to deduce the same
results about the rapidity of turbulent dissipation under likely GMC
conditions \citep{mac98,mac99,pad99}.  Similar conclusions have also
been reached concerning ongoing turbulent driving and the potential for
star formation to be initiated on a rapid timescale (see also
\cite{bal99b,elm00,kle00b}).  

In addition to studying cloud evolution, our previous work also
investigated structural properties of our model clouds.  We found that
density contrasts produced by turbulent stresses are compatible with 
the typical
clump/interclump ratio estimated in GMCs (Papers I-III).  We also found
that velocity and magnetic field power spectra evolve to be comparable
to power-law forms of Burgers and Kolmogorov turbulence, regardless of
the driving scale (Paper I; see also \cite{sto00}).  Other workers have
also studied the basic structural properties of the turbulent gas in
compressible hydrodynamic and MHD simulations, concentrating on
distribution functions of density and velocity
\citep{vaz94,pas98,sca98,nor99,kle00a}, the ability of stresses to
produce transient structure \citep{bal99a}, and power spectra and
related functions \citep{pas95,vaz97,elm97,elm99,kle00b,mac00}.

In this paper, we analyze decaying turbulence in self-gravitating
cloud models with varying mean magnetization (i.e. mass-to-magnetic
flux ratio).  \footnote{All models reported here, and most models
  studied by other workers, impose the somewhat artificial constraint
  that the initial mass-to-flux ratio is spatially uniform.}  We begin
by briefly describing the energy evolutions of our models, which serve
to confirm our earlier results on turbulent dissipation times and the
gravitational collapse criterion for magnetized clouds.  We then turn to
detailed structural investigations.
We analyze the density, velocity, and magnetic field distributions in our
models at those stages of evolution when the turbulent Mach
number is comparable to that in large ($\sim 5-10\ \pc$ scale) clouds.
Our goals are (1) to provide a basic description of structural
characteristics and how they depend on input parameters; (2) to make
connections between cloud models seen in projection and their true
three-dimensional (3D) structure, so as to help interpret
observational maps; and (3) to assess whether certain statistical
properties of clouds can be used to estimate the mean magnetization.

We divide our analysis of structure into three main sections.  The
first (\S 4) is a discussion of density structure.  Numerical
hydrodynamic and MHD simulations of supersonic, turbulent flows have
shown that magnetic pressure and ram pressure fluctuations produce
structures with large density contrast that appear to resemble
analogous ``clumpy and filamentary'' structures in real clouds
(e.g. \cite{pas95,pad99,ost99,kle00a,bal99}).  Studies of density
maxima and their immediate surroundings (``clumps'') show that many
are transient, as indicated by comparable values for the kinetic
energy and kinetic surface terms in the virial theorem (\cite{bal99a},
cf. \cite{mck92}).  Clump properties in our turbulent cloud models 
will be examined in a companion paper \citep{gam00}.  

Analysis of the {\it correlations} of overdensity -- via clump studies
or multi-point statistics -- will be needed to characterize fully how
a spectrum of self-gravitating condensations is established.  This
process is of great interest because it may ultimately determine the
stellar IMF.  A first step in understanding the effect of turbulence
on density structure, however, is to examine one-point statistics.
Here, we compute and compare the distributions of density and and
column density in different cloud models.  We consider both because
volume densities can be inferred only indirectly from observations,
whereas column density distributions can be obtained directly from
surveys of stellar extinction to background stars
(\cite{lad94,alv98,lad99}).

The second structural analysis section (\S 5) considers the linewidth-size
relation.  Observations give differing results for the slope of this
relation depending on whether the structures involved are clearly
spatially separated from the surroundings (e.g. by a large density
contrasts), or are identified as coherent regions in position-velocity
maps.  The former case yields relatively steep power spectra
\citep{lar81,sol87}; the latter case yields shallower power spectra
and larger scatter \citep{ber92,wil94,stu90}, and has led to the
concept of moderate-density ``pressure-confined clumps'' within GMCs.
We believe the different slopes are a consequence of different
definitions of ``clump.''  We use a simple binning algorithm to
explore the scaling of kinetic properties of apparent clumps within
projected clouds, and in particular to understand the consequences of
projection effects for linewidth-size relations for 2D areas and 3D
volumes.  We argue, consistent with the suggestions of some other workers
(e.g. \cite{adl92, pic00}) that it may be difficult to identify
spatially-coherent condensations from observed position-velocity maps.

The third structural analysis section (\S 6) considers the magnetic
field.  A topic of much interest in turbulence modeling is
understanding how the magnetic field affects both the
intrinsic dynamics and the observable properties of a cloud.  As
shown in \S3, a major dynamical effect of the magnetic field is to
prevent gravitational collapse in subcritical clouds.  Because
magnetic field strengths are difficult to measure directly, however,
it is highly desirable to determine if more-readily observable
structural properties of clouds could act as proxies for the magnetic
field strength. With simulations, it is possible to make 
comparisons of different models in which the mean field
strength is varied, but other key properties (such as the turbulent 
Mach number and power spectrum) are controlled.  

An important theme in 
our analysis in \S 4 and \S 5 is to test how the quantitative measures 
of structure depend on the mean magnetization, to evaluate the potential 
use of such measures as indirect magnetic field diagnostics.  In 
\S 6, we analyze how more direct magnetic field diagnostics may be 
affected by cloud turbulence.  We evaluate the distribution of total
magnetic field strength as a function of density in cloud models with
different mean magnetization.  We also compute the distribution of mean
line-of-sight integrated magnetic field (one-point statistic), which
is relevant for interpreting Zeeman effect measurements of magnetic
field strength.  Finally, we study the distribution of polarization
directions in simulated maps of polarized extinction produced by turbulent
clouds (one-point statistic).  One of the earliest estimates of
magnetic field strength in the interstellar medium (\cite{cha53}) was
based on the dispersion in polarization direction, using a simple
one-wave description of the magnetic field.  We update the
Chandrasekhar-Fermi (CF) estimate using our simulations as presumably
more realistic descriptions of the magnetic field geometry.

The plan of this paper is as follows: We start (\S 2) by describing
our numerical method and model parameters. We then (\S 3) describe our
results on energy evolution, confirming the previous results from 3D
non-self-gravitating models on dissipation rates, and 2.5D
self-gravitating models on the criterion for collapse.  We present our
structural analyses in \S\S 4-6, and conclude in \S 7 with a summary
and discussion of these investigations.

\section{Numerical method and model parameters}

We create model clouds by integrating the compressible, ideal MHD
equations using the ZEUS code \citep{sto92a,sto92b}.  ZEUS is an
operator-split, finite-difference algorithm on a staggered mesh that
uses an artificial viscosity to capture shocks.  ZEUS uses
``constrained transport'' to guarantee that ${\bf
\nabla}\cdot {\bf B}=0$ to machine precision, and the ``method of
characteristics'' to update the magnetic field in a way that ensures
accurate propagation of Alfv\'enic disturbances \citep{eva88,haw95}.  
The solutions are
obtained in a cubic box of side $L$ with grids of 256$^3$ zones, which
permits spatial resolution over a large dynamic range at manageable
computational cost. We apply periodic boundary conditions in all
models.  The simulations were run on an SGI Origin 2000 at NCSA.

For the energy equation, we adopt an isothermal equation of state with
sound speed $c_s$.  In the absence of a fully time-dependent radiative
transfer, this represents a good first approximation for the gas at
densities higher than the mean -- comprising most of the matter -- for
conditions appropriate to molecular clouds (see discussion in Paper
III, and also \cite{sca98}).  

The gravitational potential is computed
from the density using standard Fourier transform methods.  The $k =
0$ components of the density are not included in the solution due to
the periodic boundary conditions.  
Rather than the usual Poisson equation, the gravitational potential
$\phi_G$ therefore obeys $\nabla^2\phi_G = 4\pi\G (\rho-\bar\rho)$,
where $\bar\rho\equiv M/L^3$ is the mean density (mass/volume in the
box).

The initial conditions are as follows:  We start with uniform density, a
uniform magnetic field ${\bf B}_0\equiv B_0 \hat x$, and a random
velocity field $\delta {\bf v}$.  As in our earlier decay models (Papers
I-III), $\delta {\bf v}$ is a Gaussian random perturbation field with a
power spectrum $|\delta v_k|^2\propto k^{-4}$, subject to the constraint
${\bf \nabla}\cdot {\bf \delta v}=0$ so that the initial 
velocity field is non-compressive.  
This power spectrum is slightly steeper than the Kolmogorov
spectrum ($|\delta v_k|^2\propto k^{-11/3}$) and matches the amplitude
scaling of the Burgers spectrum associated with an ensemble of shocks
(but differs from Burgers turbulence in that the initial phases are
uncorrelated).

In configuration space, the velocity dispersion of the initial
conditions averaged over a volume of linear size $R$ increases as
$\sigma_v\propto R^{1/2}$.  This spectrum is comparable to the spectrum
inferred for large-scale cold interstellar clouds (e.g. \cite{lar81,
sol87, hey97})  and the spectrum that naturally arises from the
evolution of compressible turbulence that is either decaying or is
driven over a limited range of scales \citep{sto00}.  We use an
identical realization of the initial velocity field for all of the
models, so that initial states of the simulations differ only in the
strength of the (uniform) mean magnetic field.

This paper considers three different simulated cloud models.  All are
initiated with  kinetic energy $E_k=100 \bar\rho L^3 c_s^2$,
corresponding to initial Mach number ${\cal M}\equiv \sigma_v/c_s= 10
\sqrt{2}$.  For the purposes of comparison with observations, we
shall use a fiducial mean matter density 
(i.e. corresponding to the total mass divided
by total volume) $n_{H_2}=100\cm^{-3}$ and isothermal temperature
$T=10\K$ in normalizing the local simulation variables of our models to
dimensional values.
The velocity dispersion in  physical units is given by
$\sigma_v=0.19 \times {\cal M}\ \kms (T/10\K)^{1/2}$, so that the
initial value is $\sigma_v=2.7\ \kms (T/10\K)^{1/2}$.  

The models differ in their initial magnetic field strength,
parameterized by $\beta\equiv c_s^2/v_{A,0}^2= c_s^2 /(B_0^2/4\pi\bar\rho)$,
with physical value given by 
\begin{equation}
B_0=1.4\times \beta^{-1/2} \mu\G  \left({T\over 10\K} \right)^{1/2}
\left(n_{H_2}\over 100\cm^{-3} \right)^{1/2}.
\end{equation} 
We run a ``strong field'' model with $\beta=0.01$, a ``moderate
field'' model with $\beta=0.1$, and a ``weak field'' model with
$\beta=1$.  For characteristic fiducial densities and temperatures of
molecular clouds ($T\sim 10\K$, $n_{H_2}\sim 100 \cm^{-3}$), the
corresponding uniform magnetic field strengths are 14, 4.4, and 1.4 $\mu\G$.
Of course, the evolved fields are spatially nonuniform and can differ
greatly from these initial values (see \S 6), although the {\it
mean} magnetic field (i.e. the volume-averaged value or 
$k=0$ Fourier component) is a constant
$B_0 \hat x$ in time.  
The values of $\beta$ -- half the ratio of the gas pressure to
the mean field magnetic pressure --
are proportional to
the square of the mass-to-magnetic-flux ratio in the simulation box; 
this ratio cannot change in time.

We may identify several different, physically significant timescales in
the model evolution.  The sound crossing time, $t_s\equiv L/c_s$, is
fixed due to the isothermal equation of state.  Another important
timescale is the flow crossing time over the box scale $L$, $ t_f\equiv
{L\over \sigma_v} =9.8\Myr \times \left({L\over 10\pc}\right)
\left({\sigma_v \over \kms }\right)^{-1}.$ Because the turbulence decays
(i.e. $\cal M$ decreases), the instantaneous flow crossing time
increases relative to the sound crossing time as $t_f=t_s/{\cal M}$.
Where we relate $t_f$ and $t_s$, we use the Mach number associated with
the {\it initial} turbulent velocity dispersion, $\sigma_v/c_s=14.1$,
such that $t_f=0.07 t_s$.

This paper concentrates on structures that form as a consequence of
turbulence, before self-gravity becomes important.  However, we also use
the present models to test our previous results from lower-dimensional
simulations (Paper III) on the differences in the gravitational collapse
times with strong and weak mean magnetic fields $B_0$.   It is therefore
useful to define a gravitational contraction timescale
\begin{equation}
t_g\equiv \left({\pi\over \G\bar\rho}\right)^{1/2} = 9.9 \Myr 
\left({n_{H_2}\over 100\cm^{-3}}\right)^{-1/2}.
\end{equation}

In the absence of self-gravity, the unit of length $L$ defining the
linear scale of the simulation cube would be arbitrary.  In a
self-gravitating simulation, an additional parameter must be chosen to
represent the relative importance of gravity and thermal pressure
forces to the evolution.  A useful dimensionless measure of this is
$t_g/t_s$; in all the models considered here this ratio is $1/3$.  A
more transparent way of stating this is that there are three thermal
Jeans lengths $L_J\equiv c_s (\pi/G\bar\rho)^{1/2}$ across a box scale
$L$.\footnote{Of course, the presence of strong turbulence makes the
  classical, linear Jeans stability analysis inapplicable; the
  velocity field is in the nonlinear regime from the first instant.}

The three 
simulations described herein differ in the relative importance of
magnetic and gravitational forces to their ultimate evolution.
As described in Paper III, a cloud with constant mass-to-flux ratio is
super- or sub- critical if $t_g$ is smaller or larger than $\pi L/v_A$,
respectively.  A supercritical (subcritical) cloud has a ratio of
mass-to-magnetic-flux greater (smaller) than the critical value,
$1/(2\pi G^{1/2})$.  Subcritical clouds can collapse along the field but
not perpendicular to the field (``pancake''); in the nonlinear outcome
the peak density would be limited by the thermal pressure.  Supercritical
clouds can collapse both parallel and perpendicular to the field, with
unlimited asymptotic density.  The three models discussed here have
$t_g v_A/(\pi L) = 0.11, 0.34$, and $1.1$.  Thus, the strong-field
model is subcritical and the other two models are supercritical.   
The results on long-term gravitational evolution reported in \S 3 confirm the
expected differences between super- and sub- critical clouds under the
condition that turbulence secularly decays.

Since self-gravity is weak for the first portion of the evolution in
our simulations, the freedom of normalization of $L$ that applies to
non- self-gravitating models also effectively applies during this
temporal epoch.  In particular, the structural analyses of \S\S 4-6
are performed at stages of the simulations' evolutions for which the
kinetic energy is at least five times as large as the components of
the gravitational energy $E_G$ associated with the fluctuating density
distribution.  Because of our periodic boundary conditions, the
gravitational energy associated with the mean density (i.e. the $k=0$
Fourier component) is not included in $E_G$.  In order of magnitude,
the value of this lowest-order gravity is $\sim G M^2/L= M c_s^2 \pi
(L/L_J)^2$, which equals $28 M c_s^2$ for our present $L/L_J=3$
models.  This energy is in the middle of the range of kinetic energies
for the ${\cal M} =9, 7, 5$ snapshots we analyze.  Thus, as for
observed clouds (e.g. \cite{lar81, mye88}), the lowest-order
gravitational energy in these snapshots would be 
 comparable to the kinetic energy.

A useful reference length scale may be obtained by combining the
well-known observational relations between velocity dispersion, mass,
and size for GMCs (see Paper III).  The characteristic outer linear
size scale $L_{obs}$ for observed clouds scales with Mach number $\cal
M$ according to 
\begin{equation}\label{Lobs}
L_{obs}\approx 1.1\times{\cal M}\ \pc \left({T\over 10\K}\right)^{1/2} 
\left({n_{H_2}\over 100 \cm^{-3}}\right)^{-1/2}.
\end{equation}
Because the observed scale is proportional to the Mach number, the flow
crossing time for observed clouds is independent of $\cal M$, and
given by
\begin{equation}\label{tfobs}
t_{f, obs}\approx 5.8\ \Myr \times \left({n_{H_2}\over 100 \cm^{-3}}\right)^{-1/2}.
\end{equation}
For observed clouds, the flow crossing time and gravitational
contraction time are proportional, with $t_{f, obs}\approx 0.59 t_g$.  

Since the turbulence (and therefore $\cal M$) decays in our models,
they are comparable in their kinetic properties to increasingly small
clouds as time progresses.  For example, using the relation
(\ref{Lobs}), the observational scale associated with the initial
models with ${\cal M}=14.1$ would be $L_{obs,init}=16\pc\times
\left({T/ 10\K}\right)^{1/2} \left({n_{H_2}/100
    \cm^{-3}}\right)^{-1/2}$.  At this size scale, the corresponding
sound crossing time would be $t_s=82\ \Myr \times \left({n_{H_2}/ 100
    \cm^{-3}}\right)^{-1/2}$.  In the structural analyses of \S\S 4-6,
we report on properties of model snapshots in which ${\cal M} \sim 9$,
7, and 5; observed clouds of linear size scale $\sim 10,$ 8, and 6 pc,
respectively, have kinetic energies corresponding to those of the
model snapshots.  To the extent that gravity may be unimportant for much
of the internal substructure in multi-parsec scale observed clouds (suggested
by GMCs' lack of central concentration, and by the weak self-gravity of
substructures aside from the dense cores and largest clumps [e.g.
\cite{ber92,wil94}]), the correspondence between the intermediate-scale 
(``clump'') structure in real clouds and in our model snapshots may be 
quite direct.  

Since some ambiguity remains in associating an overall physical length
scale with our simulated models (due to the periodic boundary
conditions), we report integrated quantities solely in dimensionless
units, giving e.g. column densities in units of the mean column
density, $\bar N\equiv \bar\rho L$.  For local variables (such as
magnetic field strengths) which bear no such ambiguity, we report
values in dimensionless units and also transform to physical units
based on our adopted fiducial density and temperature.

\section{Energy evolution in model clouds}

The early evolution in all the models follows a similar course.
Kinetic energy initially
decreases as the fluid works to deform the magnetic field.  The
initially non-compressive velocity field is transformed into a compressive
field, by interactions with the magnetic field and nonlinear coupling of
the spatial Fourier components.  This leads to the
development of density-enhanced and density-deficient regions, 
and results in the dissipation of energy in shocks.  
Fluctuations in the density cause fluctuations in the
gravitational potential that begin to dominate the dynamics at late
times, and lead to runaway gravitational collapse for supercritical 
models.

\def\d3r{d^3{\bf r}\,}

To quantify the energetic evolution, we define the kinetic energy
\begin{equation}\label{ekdef}
E_K={1\over 2} \int \d3r
(v_x^2 + v_y^2 + v_z^2)\rho,
\end{equation}
the perturbed magnetic energy
\begin{equation}\label{ebdef}
\delta E_B={1\over 8\pi} \int \d3r
(B_x^2+B_y^2+B_z^2) - E_{B,0},
\end{equation}
where $E_{B,0}= L^3 B_0^2/8\pi$ is the energy in the mean magnetic
field, and the gravitational potential energy
\begin{equation}
E_G={1\over 2} \int \d3r (\rho \phi_G),
\end{equation}
where $\phi_G$ is the gravitational potential computed from the Poisson
equation modified for periodic boundary conditions.  

\begin{figure}
\plotone{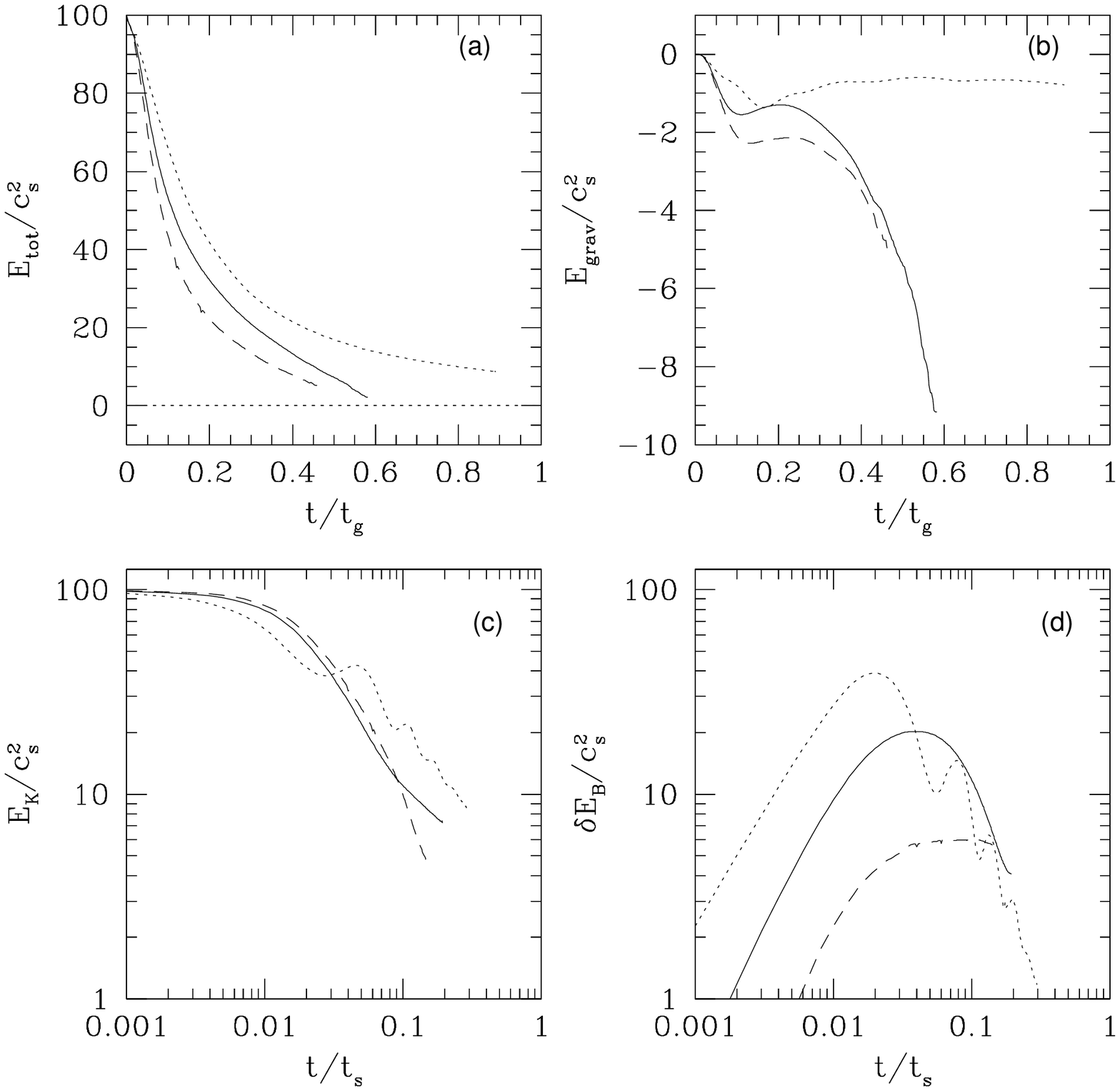}
\caption{Energy evolution of model clouds. Models with $\beta=0.01$, $0.1$, 
and $1$ are shown with dotted, solid, and dashed curves, respectively.
\label{fig-enev}}
\end{figure}

Figure \ref{fig-enev} shows the evolution of $E_{tot} \equiv
E_K+\delta E_B+E_G$, $E_G$, $E_K$, and $E_B$ (see panels $a$-$d$,
respectively).  From Fig. \ref{fig-enev}b, it is clear that,
consistent with expectations and previous results on self-gravitating
cloud models with decaying turbulence (Paper III), all but the
magnetically subcritical $\beta=0.01$ model suffers a gravitational
runaway.  Both of the supercritical models become gravitationally
bound at time $\sim 0.6 t_g,$ corresponding to $\sim 6 \, \Myr
\left({n_{H_2}/{10^2\cm^{-3}}} \right)^{-1/2}$.  \footnote{The
  simulations are terminated shortly after the onset of gravitational
  runaway because the coincident development of low-density regions
  where $v_A$ is large causes the Courant-condition-limited timestep
  to become very short.  For the $\beta=1$ model, the gravitational
  binding time (the time to reach $E_{tot}=0$) is an extrapolation
  based on the evolutions of the $\beta=0.1$ and $\beta=1$ models
  shown in Figs.  \ref{fig-enev}a,b.} The gravitational runaway time
is comparable to that found in lower-dimensional simulations.

The kinetic energies in all models decay rapidly.  After one flow
crossing time $t_f$, the kinetic energy has been reduced by $73-85\%$
compared to the initial value (see Table \ref{tbl-1}).  The kinetic
energy is reduced by a factor two after $0.2-0.4$ flow crossing times
(Table \ref{tbl-1}), with this kinetic loss time decreasing toward
lower $\beta$ (stronger $B_0$) because of the more rapid transfer of
kinetic to perturbed magnetic energy when the Alfv\'en frequency is
higher.  For GMC parameters (see eq. \ref{tfobs}), the corresponding
dimensional kinetic energy decay time would be 1-2 Myr.  The growth of
magnetic energy stored in these magnetic field fluctuations (due to
advection by the turbulent velocity field) is apparent in Figure
\ref{fig-enev}d; the initial increase is followed by decreasing or flat
perturbed magnetic energy as the turbulent velocity field decays.  The
time to reach the maximum perturbed magnetic energy lies in the range
$0.1-0.2$ times the Alfv\'en crossing time, similar to what was found
in lower-dimensional simulations (Paper III).  At the point when
$\delta E_B$ is maximal, it accounts for 20-50\% of $E_{turb}$.  The
fraction increases with the mean field strength $B_0$.

\begin{figure}
\epsscale{.5}
\plotone{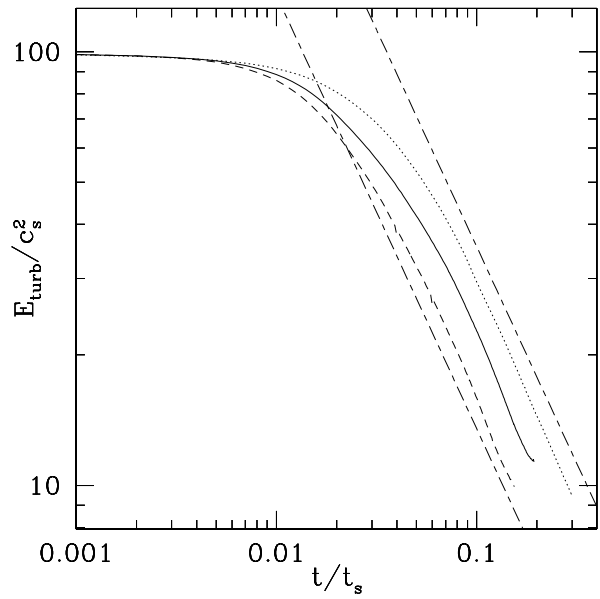}
\caption{Evolution of turbulent energy $E_{turb}=E_K+\delta E_B$ in 
model clouds. Models with $\beta=0.01$, $0.1$, 
and $1$ are shown with dotted, solid, and dashed curves, respectively.
The dot-dash lines indicate a slope $\propto t^{-1}$, for reference.
\label{fig-eturb}}
\end{figure}

The total ``turbulent'' energy ($E_{turb}=E_K+\delta E_B$) secularly
decreases in time; after somewhat more than half of the initial
turbulent energy is lost, the decay approaches a power-law temporal
behavior with $E_{turb}\propto t^{-1}$ (Figure \ref{fig-eturb}).  This
late-time scaling in non-self-gravitating models of 3D MHD turbulence
has been noted previously \citep{mac98,sto98,mac99}.  Most of the
turbulent losses, however, occur before the onset of this behavior.  The
turbulent decay can be characterized by the time $t_{dec}$ to reduce the
turbulent energy by a factor two from its initial value.  We find that
this time is in the range $0.4-0.8$ flow crossing times (see Table
\ref{tbl-1}), comparable to our results from Paper II, and consistent
with other findings (Paper III, \cite{mac98,mac99},\cite{pad99}) that
dissipation times vary by only a factor $\sim 2$ over the range of Mach
numbers and magnetic field strengths present in GMCs.  The corresponding
dimensional time for turbulent energy decay with GMC parameters is
2-4 Myr.

\textwidth=6.5in


\begin{deluxetable}{crccccc}
\tabletypesize{\small}
\tablewidth{5in}
\tablecaption{Comparative Energy Evolution\label{tbl-1}}
\tablehead{
\colhead{model}
& \colhead{$\beta$} 
& \colhead{$E_{K}(t_f)/E_{\rm K, init}$\tablenotemark{a}} 
& \colhead{$\delta E_{B}/E_{K}(t_f)$\tablenotemark{a}} 
& \colhead{$t_{dec}/t_f$\tablenotemark{a,b}} 
& \colhead{$t_{dec}^K/t_f$\tablenotemark{a,b}} 
& \colhead{$t_{bind}/t_g$\tablenotemark{c}} 

}
\startdata
 B&  $0.01$&  $0.27$&  $0.51$& $0.76$    & $0.21$   & $>0.9$\\
 C&   $0.1$&  $0.15$&  $1.13$& $0.55$    & $0.31$   & $0.6$\\
 D&   $1.0$&  $0.17$&  $0.35$& $0.41$    & $0.37$   & $0.6$\\ 
\enddata
\tablenotetext{a}{The flow crossing time 
$t_f\equiv L/(2 E_{\rm K, init})^{1/2}\rightarrow 0.07 t_s$}
\tablenotetext{b}{$t_{dec}$ ($t_{dec}^K$)  is the time to 
reduce the initial energy (kinetic energy) by 50\% }
\tablenotetext{c}{$t_{bind}$ is the time at which 
$E_K+\delta E_B +E_G=0$; $t_g\equiv (\pi/G\bar\rho)^{1/2}\rightarrow 0.33 t_s$}
\end{deluxetable}


\section{Density and column density distributions}

A basic statistical property of a real or model cloud is the
distribution of density in its constituent parts.  This distribution
may be described either by its fractional volume per unit density
($dV/d\rho$) or by its fractional mass per unit density ($dM/d\rho$).
Previous analyses of the density distributions in compressible
hydrodynamic turbulence simulations (before gravity becomes important)
show that when the equation of state is approximately isothermal, the
density distribution is close to a log-normal
\citep{vaz94,pad97,pas98,ost99}.  \cite{sca98}, \cite{pas98}, and
\cite{nor99} also show that in a medium where the temperature decreases
(increases) with increasing density, an extended tail in the density
distribution function develops at density higher (lower) than the mean
density.  In the present models we assume an isothermal equation of
state.  This is is a reasonable approximation here since most of the
gas is contained in condensations at density larger than the mean
value where the temperature likely varies by less than a factor $\sim
2$ (see Paper III, also \cite{sca98}).

As described in \S 1, a key question is whether it is possible to
discriminate the magnetic field strength in a cloud from its
structural properties.  Using our present models, we can test how the
strength of the mean magnetic field affects the observable
density and column density statistics.  For these tests, we choose
sets of model ``snapshots'' from the three decay models 
in which the Mach number ${\cal M}\equiv \left<v^2/c_s^2\right>^{1/2}$ 
(or kinetic energy) matches in the three models;  
because the energy evolves at somewhat different rates in the 
runs with different $\beta$, these times of the snapshots vary. The sets
of model snapshots have ${\cal M}\approx 9, 7, 5$.

Figure \ref{fig-rhohist} shows an example of the distributions of volume
and mass as a function of volume density, $\rho$, where $\rho$ is
measured in units of the mean density $\bar\rho=M/L^3$ in the simulation
cube.  The volume density distributions are well approximated by
log-normal functions, i.e., volume and mass distributions in $y\equiv
\log(\rho/\bar\rho)$ of the form 
\begin{equation}\label{eq-logndef}
f_{V,M}(y)={1\over \sqrt{2\pi \sigma^2}} \exp[-(y\pm|\mu|)^2/2\sigma^2],
\end{equation}
where the upper/lower sign on the subscript applies to the volume/mass
distribution; $f_{V,M} dy$ is the fraction of the volume or mass with
$y$ in the interval $(y,y+dy)$.  It is straightforward to show that the
mean $\mu$ and dispersion $\sigma$ are related by
$|\mu[y]|=\ln(10) \sigma[y]^2/2$ for a log-normal distribution, so
$\sigma[y]=0.93\sqrt{\mu[y]}$.\footnote{Elsewhere, distributions are
sometimes given as a function of $x\equiv \ln(\rho/\bar\rho)$; in that
case, $|\mu[x]|=\sigma[x]^2/2$ since $\mu[y]=\mu[x]/\ln(10)$ and
$\sigma[y]=\sigma[x]/\ln(10)$.}  Table \ref{tbl-2} (cols. 5,6) gives the
values of $\mu_V[y]$ ($\equiv\langle y \rangle_V$) 
and $\mu_M[y]$ ($\equiv\langle
y\rangle_M$) for three sets of models at different Mach numbers (where
the subscript on the angle brackets denotes weighting by volume or
mass). In all cases, $\mu_V[y]\sim-\mu_M[y]$, consistent with a
lognormal distribution.



\begin{deluxetable}{crccccccc}
\tabletypesize{\small}
\tablewidth{5in}
\tablecaption{Comparative Density and Column Density \label{tbl-2}}
\tablehead{
\colhead{snapshot}
& \colhead{$\beta$} 
& \colhead{$t/t_s$}
& \colhead{$\cal M$}
& \colhead{$\mu_V[y]$\tablenotemark{a}}
& \colhead{$\mu_M[y]$\tablenotemark{a}}
& \colhead{$\mu_{M;x}[Y]$\tablenotemark{b}}
& \colhead{$\mu_{M;y}[Y]$\tablenotemark{b}}
& \colhead{$\mu_{M;z}[Y]$\tablenotemark{b}}
}
\startdata
 B1&  $0.01$&  $0.03$&  $8.8$& $-0.25$ & $0.30$ & $0.024$ & $0.033$ & $0.029$\\
 C1&   $0.1$&  $0.03$&  $8.8$& $-0.27$ & $0.28$ & $0.030$ & $0.037$ & $0.047$\\
 D1&   $1.0$&  $0.03$&  $9.4$& $-0.42$ & $0.37$ & $0.044$ & $0.047$ & $0.061$\\
 B2&  $0.01$&  $0.07$&  $7.4$& $-0.38$ & $0.38$ & $0.038$ & $0.037$ & $0.046$\\
 C2&   $0.1$&  $0.04$&  $7.6$& $-0.27$ & $0.29$ & $0.034$ & $0.046$ & $0.048$\\
 D2&   $1.0$&  $0.05$&  $7.2$& $-0.37$ & $0.34$ & $0.060$ & $0.054$ & $0.065$\\
 B3&  $0.01$&  $0.19$&  $4.9$& $-0.23$ & $0.21$ & $0.015$ & $0.028$ & $0.021$\\
 C3&   $0.1$&  $0.09$&  $4.9$& $-0.35$ & $0.37$ & $0.047$ & $0.050$ & $0.056$\\
 D3&   $1.0$&  $0.09$&  $4.9$& $-0.31$ & $0.33$ & $0.048$ & $0.057$ & $0.063$\\
\enddata
\tablenotetext{a}{Volume-weighted or mass-weighted 
average of the logarithmic density contrast, 
$y\equiv\log(\rho/\bar\rho)$; expected 
sampling error is $\sim 10^{-4}$.}
\tablenotetext{b}{Mass-weighted average of logarithmic column density 
contrast, $Y\equiv \log(N/\bar\rho L)$, for 
projection along $\hat x$, $\hat y$, or $\hat z$;
expected sampling error is $\sim 10^{-3}$}
\end{deluxetable}


For a log-normal distribution, the weighted mean and dispersion of the density
itself are related to the mean of the logarithmic density contrast 
$\mu[y]$ using
\begin{equation}\label{eq-lognmn}
\log\left<{\rho\over\bar\rho}\right>_M =  
2\left<{\log{\rho\over \bar\rho}  } \right>_M \equiv 2|\mu[y]|,
\end{equation}
\begin{equation}\label{eq-logndv}
\left<\left({\rho\over\bar\rho}\right)^2\right>_V=
\left<{ \rho\over \bar\rho}\right>_M=10^{2|\mu[y]|},
\end{equation}
and 
\begin{equation}\label{eq-logndm}
\left<\left({\rho\over\bar\rho}\right)^2\right>_M=
\left<{\rho\over\bar\rho}\right>_M^3=10^{6|\mu[y]|},
\end{equation}
where ``V'' and ``M'' subscripts denote weighting by volume and mass,
and $\langle \rho\rangle_V\equiv \bar\rho$.  From Table \ref{tbl-2},
$\mu[y]$ is in the range $\sim 0.2 - 0.4$ for ${\cal M}=5-9$, implying
from equation (\ref{eq-lognmn}) that the typical mass element has been
compressed by a factor 
$\langle \rho/\bar\rho\rangle_M \sim 2.5-6$ compared to its
unperturbed initial value.  Because of the log-normal form of the
distribution, two-thirds of the matter is within a factor
$10^{0.93\sqrt|\mu[y]|}$ ($\sim 2.7-3.8$) above or below the value
$10^{|\mu[y]|}\bar\rho\sim (1.6-2.5)\bar\rho$, and 95\% is within a
factor $10^{1.86\sqrt|\mu[y]|}$ ($\sim 7.2- 14$) above or below this value.
The volume-weighted rms standard deviation in $\rho/\bar\rho$ is 
$(10^{2\mu[y]}-1)^{1/2}$ ($\sim 1.3-2.2$), and the 
mass-weighted rms standard deviation in $\rho/\bar\rho$ is 
$10^{2\mu[y]}(10^{2\mu[y]}-1)^{1/2}$ ($\sim 3-13$).

The above results on density contrast may be compared with previous
work.  In Paper III, we found that for 2.5 dimensional models of
decaying turbulence with $\beta=0.01,0.1,1.0$, the mean logarithmic
density contrast $|\mu[y]|=0.2-0.5$ for Mach numbers in the range
$5-10$, with a weak trend toward an increase in the contrast with
increasing Mach number, and the largest contrast in the strong-field
($\beta=0.01$) group.  For the quasi-steady forced-turbulence models
with ${\cal M}\approx 5$ reported on in Paper II, the mean logarithmic
density contrasts $|\mu[y]|$ are in the range 0.20-0.28, increasing from
$\beta=1.0$ to $0.01$.  Thus, overall, we find comparable values of the
density contrast in all our analyses of turbulence in those stages where
self-gravity is not important.  

\cite{nor99} and \cite{pad97} report findings implying that, for 3D
unmagnetized forced turbulence, $\mu[y]$ is related to the Mach number
$\cal M$ by $|\mu[y]| = (1/2)\log(1+ 0.25{\cal M}^2)$.  For the range of
Mach numbers (${\cal M}\sim 5-9$) in our Table \ref{tbl-2}, the
corresponding values of $|\mu[y]|$ would be $\sim 0.4-0.7$, somewhat
larger than those we found; however, \cite{nor99} remark that they find
lower density contrasts when the magnetic field is nonzero, which would
yield better agreement with our results for magnetized turbulence.
Analysis of simulations of compressible, isothermal, unmagnetized
turbulence in one dimension by \cite{pas98} suggest a linear rather than
logarithmic scaling for $\mu[y]$ with Mach number $\cal M$ and much
larger values of the contrast than those found in 3D simulations.  This
may be due to the purely compressive velocity field in 1D.

The present analysis suggests that for non-steady magnetized turbulence,
there is no one-to-one relationship between the density contrast and the
sonic Mach number or other simple characteristic of the flow.  There
does, however, appear to be a secular increase in the minimum value of
the contrast with the effective Mach number for magnetized flow,
the fast magnetosonic Mach number ${\cal M}_F$ defined by ${\cal
M}_F^2\equiv \langle v^2\rangle/\langle v_{A}^2 +c_s^2\rangle$.  
In Figure \ref{fig-rho_vs_mach}, we
plot the logarithmic contrast factors against the value of $\log(1+{\cal
M}_F^2)$; the lower envelope of the contrast is found to be fit by
$|\mu[y]|=0.2 (\log(1+ {\cal M}_F^2) +1)$ for the models studied.  There
is no similar secular relationship between the density contrast and the
ordinary sonic Mach number $\cal M$.  

Applying similar reasoning to the argument of \cite{pas98}, the weak
relation between the effective Mach number ${\cal M}_F$ and the density
contrast may be understood heuristically as follows.  From equations
(\ref{eq-lognmn}) and (\ref{eq-logndm}), we may write the mean
logarithmic contrast in terms of the mass-weighted dispersion in density
amplitude as 
\begin{equation}\label{eq-lognsolve}
\left<\log\left({\rho\over\bar\rho}\right)\right>_M = {1\over 6} 
\log\left[\left<\left({\rho\over\bar\rho}\right)^2\right>_M \right].
\end{equation}

In strong, unmagnetized, isothermal shocks, which would occur for flow
parallel to the field, the preshock and postshock densities $\rho_1$
and $\rho_2$ have $(\rho_2/\rho_1) -1\approx {\cal M}^2$.  For strong
isothermal shocks magnetized parallel to the shock front and
$\beta\simlt 1$, $(\rho_2/\rho_1) -1$ is linear rather than quadratic
in ${\cal M}_F$, approaching $\sqrt{2}v/v_A\sim \sqrt{2}{\cal M}_F$.
If the typical shock jump compression factor determines the rms
dispersion in the density, then the term in square brackets in
equation (\ref{eq-lognsolve}) would scale between quadratically and
quartically in ${\cal M}_F$ for a range of $\beta$ and shock
geometries (noting that ${\cal M}_F\rightarrow {\cal M}$ for $\beta$
large).  The real situation is of course more complicated.  It is
interesting, however, that the slope $\sim 0.2$ of the lower envelope
of the $<\log(\rho/\bar\rho)>$ vs. $\log (1+{\cal M}_F^2)$ relation
does fall in the range between $0.17$ and $0.33$ suggested by this
heuristic argument.  The fact that this lower envelope lies closer to
the (shallower) slope corresponding to parallel-magnetized shocks
indicates that the model turbulent clouds do not invariably evolve to
be dominated by (more compressive) flows aligned with the mean
magnetic field.

Because of the potential for direct comparison with observation, it
useful to examine the distributions of column density $N$.   In
particular, we would like to ascertain if the distribution depends on
the mean magnetization.  The distribution of column densities can be
described by the fractional area, $dA/dN({\hat s})$, or fractional mass,
$dM/dN({\hat s})$ per unit column density, where ${\hat s}$ is the
orientation of the line of sight through the cloud.  In Figure
\ref{fig-colhist}, we compare the distributions of projected area and
mass as a function of column density for model snapshots (B2,C2,D2 from
Table \ref{tbl-2}) with matched Mach numbers and different values of the
mean magnetic field strength.  
Although the statistics are poorer than for the distributions of volume
density, the column density distributions are also approximately
log-normal in shape.  Thus, the column density distributions can be
described using the same form as equation (\ref{eq-logndef}), but
replacing $y\rightarrow Y\equiv \log(N/\bar N)$, where $\bar N \equiv
\bar\rho L$.  The mean and dispersion may depend on the projection
direction $\hat s$, so $\mu[y]\rightarrow \mu_s[Y]$, and $\sigma[y]\rightarrow
\sigma_s[Y]$.  To the extent that the distributions follow log-normal
forms, the formal relations (\ref{eq-lognmn})-(\ref{eq-logndm}) would
apply, with $\rho/\bar\rho\rightarrow N/\bar N$ and area-weighting
replacing volume-weighting.  

In Table \ref{tbl-2}, we list the values of the mass-weighted mean of
the logarithmic column density contrast ($|\mu_{M;s}[Y]|\equiv
\left<\log(N/\bar N) \right>_{M;s}$) for the different model snapshots
in each of three projection orientations $\hat s=\hat x,\hat y, \hat z$.
From the data in the Table, the projections in the various directions
for any model snapshot yield somewhat different statistics (mostly
10-20 \% differences in $\sigma[Y]$ and twice that in $\mu[Y]$);
the projection along the magnetic field tends to give slightly
lower contrast than the two perpendicular projections.  
Differences between the two perpendicular directions ($\hat
z$ and $\hat y$) are simply a result of specific realizations of random
initial conditions.  Because the models have the same initial
velocity perturbation realization, they will have similar evolved
structure to the extent that the magnetic fields only weakly affect the
dynamics -- this explains, for example, why models C and D both have
larger contrast for $\hat z$ projections than $\hat y$ projections.

Notice that models with the strongest magnetic field tend to have lower
column density contrasts than models with the same Mach number and
weaker mean $B_0$ (20-50\% differences in $\sigma[Y]$ for most
sets).  This effect is most pronounced for the Mach-5 set (B3, C3, D3;
this set has a factor two [three] difference in the $\sigma[Y]$
[$\mu[Y]$]).  \cite{pad99} previously pointed out that column density 
contrasts may be larger in weaker mean-$B_0$ models. Our results 
confirm this tendency, although we find that the effect is relatively
weak in magnitude, and does not hold in all cases (see e.g. the 
results for snapshots B2 and C2 in the Table).

Overall, the range of mean logarithmic column density contrasts in
Table \ref{tbl-2} is $\mu[Y] \sim 0.015-0.065$, corresponding to
typical mass-averaged column density in the range $\langle N/\bar
N\rangle_M = 1.07-1.35$, i.e., only a modest enhancement over the
average in a uniform cloud.  The range of logarithmic column density
contrasts is an order of magnitude lower than the range of mean
logarithmic density contrasts.  This is understandable, since each
column contributing to the distribution samples a large number of
over- and under- densities along the line of sight.  The column
density distributions still require a density correlation length over
a significant fraction of the box size $L$ along the line of sight,
however; otherwise the dispersion in column densities would be wiped
out by line-of-sight averaging.

This can be seen more quantitatively as follows.  Each of $n_A$ columns
that contributes to the distribution is created by taking the sum of
densities in $n_s$ cells along the line of sight.  From the Central
Limit Theorem, we know that {\it if} the density in each cell along the
line-of-sight {\it were}  an independent random variable, then for $n_s$
large, the distribution of column densities would approach a {\it Gaussian} 
-- rather than log-normal -- shape, with (area-weighted) mean of $N/\bar N$
equal to 1 (where $\bar N\equiv \bar\rho L$), and (area-weighted) 
standard deviation in $N/\bar N$ equal to
$n_s^{-1/2}$ times the (volume-weighted) standard deviation 
in $\rho/\bar\rho$.  For a log-normal volume
density distribution obeying equation 
(\ref{eq-logndv}), an assumption of independent sampling along the 
line-of-sight would therefore predict an (area-weighted) 
rms deviation of $N/\bar N$ from unity given by
\begin{equation}\label{eq-gausscol}
\sigma_{N/\bar N}^{Gauss} = {1\over \sqrt{n_s}}\left(10^{2|\mu[y]|} -1\right)^{1/2},
\end{equation}
with typical sampling error $\sim \sigma_{N/\bar N} /\sqrt{n_A}$ in
determining the mean and dispersion of $N/\bar N$.  For $|\mu[y]|\sim
0.2 -0.4$, the expected standard deviation in $N/\bar N$ would be $\sim 0.08
-0.14$, with sampling error $\sim 0.0003-0.0006$, if the line-of-sight 
cells were all independent.  {\it In fact}, using
the area-weighted equivalent of relation (\ref{eq-logndv}) for
the log-normal ({\it not} Gaussian) {\it column} density distribution
that is evidently produced, 
the area-weighted standard deviation in $N/\bar N$ is 
\begin{equation}\label{eq-logncol}
\sigma_{N/\bar N}^{log-norm} = \left(10^{2|\mu[Y]|} -1\right)^{1/2},
\end{equation}
or approximately $\sqrt{2\ln(10)\mu[Y]}$ for $\mu[Y]<<1$. For our
tabulated values, this is in the range
$0.27-0.59$, significantly larger (by hundreds of times the sampling
error) than would be predicted by assuming uncorrelated values of the
density along any given line of sight.  Thus, both the non-Gaussian
shape and the breadth of the dispersion of the column density distributions
argues that the volume
densities are {\it not} independent but are correlated
along any line of sight -- as indeed should be expected since there are
large coherent regions of density created by the dynamical flow.  

We speculate that it may be possible to understand the dynamical
process behind the development of the log-normal {\it column} density
distribution following similar reasoning to the argument of
\cite{pas98} for the development of a log-normal {\it volume} density
distribution.  They argue that if consecutive local density
enhancements and decrements occur with independent multiplicative
factors due to independent consecutive velocity compressions and
rarefactions, then the log of the density in some position is the sum
of logs of independent enhancement/decrement factors; this would yield
a lognormal density distribution if there are many independent
compressions/rarefactions, each sampling independently from the same
distribution of enhancment/decrement factors.

Suppose, similarly, that the gas along any line of sight is 
subject to multiple independent compression/rarefaction events;  since the
compression/rarefaction axes are not in general along the line-of-sight,
column density on a given line-of-sight is not conserved.  Each 
compression/rarefaction event which produces a local change in the
volume density by a factor $X$ affects only a fraction $f$ of the column
of gas, resulting in an {\it effective} enhancement/decrement factor for the
column closer to unity than $X$.  
A simple model would be to suppose that each event
$i$ independently produces a change in the column density by a factor
$X_i'=(1-f_i) + f_i X_i= X_i -(1-f_i)(X_i-1)$ 
(taking the fraction $1-f_i$ of the gas in the column at
unchanged volume density and the fraction $f_i$ at volume density 
enhanced/decreased by a factor $X_i$).  If $X_i >1$ (respectively, $X_i<1$), then
$X_i'<X_i$ (respectively $X_i'>X_i$).  The logarithm of the column density
contrast would then be a sum of terms $\log X_i'$;  taking these 
as random variables, the resulting distribution would be log-normal
(assuming a large number of [spatially overlapping] successive events).  
Since each $X_i'$ is closer to unity than $X_i$, the mean and dispersion
of the logarithmic column density distribution are expected to be smaller
than those of the volume density distribution.  
Although it would be interesting to test in
detail whether this sort of heuristic model could be refined and 
used to relate
projected density distributions to volume density distributions, the
potential for finding a unique inversion (even in a statistical sense)
is limited by the many degrees of freedom associated with the extended
spatial power spectrum producing the compressions.

For the power-law input turbulent spectrum that we adopt, the spatial
correlations which produce the column density distribution occur at
sufficiently large scale that the distributions are not, except at
columns $N$ much larger than the mean, very sensitive to the
``observer's'' resolution.  For example, Figure \ref{fig-colres} shows
the statistics of column density for one model at the full resolution of
the simulation, and for resolving power reduced by factor of four by
averaging the column density values within squares of edge size four
times that of simulation cells (so that each ``pixel'' has sixteen times
the area of a projected simulation cell).  The overall shapes and mean
values of the distributions are quite comparable.  At column densities
much larger than the mean, of course, the distributions become sensitive
to resolution because of the scarcity of regions with the highest column
density; averaging these with their lower-column-density neighbors
results in a cutoff of the distribution at lower $N$.  A related point
for observed $^{13}$CO data was discussed by \cite{bli97}.  They showed
that the distribution of the number of cells in position-velocity space
as a function of $T_A/T_{A,max}$ in the cell  flattens as the linear
resolution scale increases, due to the smearing-out of the
highest-column regions.   We have verified that the distribution of
number of projected cells with $N/N_{max}$ similarly becomes flatter if
the map of projected density is averaged over grids with increasing cell
size. 

Because the periodic boundary conditions introduce an effective
correlation in the density along the line-of-sight at scale $\sim L$,
a potential concern might be that the typical column density contrast
might be enhanced by introducing ``artificial'' coherence along lines
of sight.\footnote{We thank E. V\'azquez-Semadeni for noting this point.}  
To investigate this effect, we have evaluated sets of
``half-column'' density distributions by summing only over distances
$L/2$ along the line-of-sight.  In general, the resultant half-column
distributions are still lognormal in form (although noisier), with
larger means and dispersions than those found for the full-column
integrations.  Figure \ref{fig-colhistfront} shows one such set of
distributions, obtained from the $z>L/2$ ``front half'' of the volume
snapshots B2, C2, D2.  This result suggests that the coherent volume
density regions responsible for the lognormal column density
distribution in fact have intermediate scale -- they are much larger
than the cell size, but significantly smaller than the overall size of
the box.  In this situation, one would expect that a factor two
decrease in the number of (multi-cell) correlated regions along the
line-of-sight would produce a factor $\sqrt{2}$ increase in
$\sigma_{N/\bar N}$, corresponding to a factor $\sim 2$ increase in
$|\mu[Y]|$ (cf. eq. \ref{eq-logncol}).  
Indeed, we find the half-column values of $|\mu[Y]|$ are
typically larger than the full-column values by a factor $\sim 1.5-2$,
supporting this interpretation.

The robustness of the column density distribution to resolution
changes makes it a viable statistic for comparing simulations to the
observable properties of turbulent clouds.  Such comparisons are a
test of the idea that much of the moderate-density ``clumpy''
structure in molecular clouds may be produced by turbulent stresses.
Preliminary results are promising; for example, we have compared the
distribution of the extinction data values from the dark cloud IC5146
\citep{lad99} with column density distributions from our simulation
snapshots.  Figure \ref{fig-cumcol} shows that the cumulative
distributions are indeed remarkably similar in form (although this 
particular real cloud has a slightly larger dispersion than our models have).  
Unfortunately, however, the
column density distribution is determined by more than just a few
simple global parameters.  In some circumstances, there may be as much
variation in the column density distributions between the same cloud
viewed at different orientations as there is in two clouds with the
same turbulent Mach number but a factor ten difference in the mean
magnetic field.  This large ``cosmic variance'', and the relatively
weak variation with parameters of $\mu$ and $\sigma$ compared to their
scatter, make it unlikely that it will be possible to estimate
individual clouds' mean magnetic field strengths, for example, from
column density distributions alone.

\section{Linewidth-size relations and projection effects}

An important way of characterizing the kinetic structure in turbulent
clouds is to measure the distribution of the velocity dispersion {\it
vs.}  the physical size or mass of the regions over which it is
averaged.  Means over these distributions then represent
``linewidth-size'' relations.  The regions over which velocity
dispersions are averaged in observed clouds are often apparent
``clumps''.  At the most basic level, an apparent ``clump'' in a cloud
or projected cloud is a spatially connected, compact region that
stands out against the surrounding background.  In any hierarchical
structure, clumps will contain other smaller clumps, and in general
the identification of clumps is a resolution-dependent procedure.
Starting from the fundamental concept of a ``clump'' as a region of
contrast (ROC) on a given spatial scale, we have developed a simple
algorithm to identify and characterize the ensemble of projected ROCs
at multiple scales, so as to explore the scaling of kinetic properties
with physical size.\footnote{In \cite{gam00}, we use an alternative approach
to define clumps and characterize their properties.}

The procedure is as follows: First,
we choose a size scale $s$ (here, a factor $2^n$ times the simulation
grid scale, where $n=2-8$).  We then divide the projected cloud into
zones of area $s^2$.  Within each zone, we compute the mean projected
surface density as the total zone mass divided by $s^2$; we also compute
the mass-weighted mean surface density $\Sigma[s]$ for the set of zones
on scale $s$.  We label a
zone as a ROC on scale $s$ if its surface density is at least a factor
$f_c$ times $\Sigma[s]$.  Typically we use $f_c=1$, but the results
are not qualitatively sensitive to this choice; we note that (i) regions
above the mean column density at a given scale occupy less than half 
the area due to mass conservation, and (ii) 
since $\Sigma[s]$ increases with decreasing $s$, the ROCs on a given scale
would appear ``by eye'' to stand out against the background
even with $f_c=1$.  For each projected
ROC, we also compute the (mass-weighted) dispersion of the line-of-sight
velocity $\sigma_v$; this represents the ``linewidth'' for a region of
projected area $s^2$.  

We are now in a position to examine the correlations among linewidth
$\sigma_v$, mass $M$, and spatial size $s$ for our ROC collections.  In
a data set based on molecular line emission, the contributions from any
local region would depend on the local excitation rather than
simply being proportional to the amount of matter present.  For the
analysis described below (except as noted), 
we only include contributions from material if
its local density (mass/volume) is at least equal to 
$\rho_{min}=3\bar\rho$, as a
simple way of selecting material in the range of densities that contribute 
to common molecular lines.\footnote{Realistically, of course, the 
contribution to observed lines depends on more than the local 
density;  due to radiative transfer effects, it might even be possible for
lower-density material to contribute more efficiently than higher-density
material if its emission occurs in line wings and suffers less absorption.}

Figure (\ref{fig-roc}) shows an example of how the ROCs at multiple
scales are distributed on the map of model snapshot B2 projected in
the $\hat z$ direction (Figure \ref{fig-imagelb} shows a colorscale
image of the column density for the same snapshot projection).  For
the ROC ensemble shown in the figure, we compute masses, velocity
dispersions, and values of the so-called virial parameter $\alpha
\equiv 5 \sigma_v^2 s/G M$ \cite{ber92}.  In Figure (\ref{fig-lws}),
we plot the values as a function of (linear) size scale and/or mass.
We also evaluate least-squares linear fits to
$d\log(\sigma_v)/d\log(s)$, $d\log(M)/d\log(s)$,
$d\log(\sigma_v)/d\log(M)$, and $d\log(\alpha)/d\log(s)$; the
respective values in this example are $0.09$, $1.87$, $0.06$, and
$-0.4$.

From parts (a) and (c) of Figure (\ref{fig-lws}), it is clear that
although the there is a {\it mean} increase in velocity dispersion with
mass and linear size, there is a great deal of scatter as well.  The
upper envelopes of the velocity dispersion distributions in fact even
{\it decrease} as a function of increasing $M$ and $s$;  the lower
envelopes  increase more steeply.  The distribution of $\alpha$ vs. $M$
also shows large dispersion, with a nearly-flat lower envelope and an
upper envelope showing a decrease in $\alpha$ with $M$.  The $M$ vs.
$s$ distribution has a relatively low dispersion.   

Many of the features evident in Figure (\ref{fig-lws}) can be understood
by reference to the scaling properties of the underlying
three-dimensional distribution, together with the effects of projection
onto a plane.   First consider the $\log(\sigma_v)$ - $\log(s)$
distribution.  The procedure we have used to identify ROCs in the
projected plane also can be used in the 3D data cube itself; we can then
compute the mass and velocity dispersion for each 3D cell of edge size
$s$ that meets the contrast criterion.  In Figure (\ref{fig-lws} a),
we show how the mean velocity dispersion for these 3D cells depends on
size scale.  Interestingly, this curve traces fairly closely the {\it
lower envelope} of the distribution of $\sigma_v$ vs. projected size for
ROCs on the projected plane.  Thus, for nearly all projected regions of
area $s\times s$, the majority the velocity dispersion can be attributed
to the superposition along the line-of-sight of many regions of volume
$s\times s\times s$  with different mean velocities.  The relatively
weak dependence of mean linewidth on {\it projected} size (or mass)
simply reflects the ubiquity of ``contamination'' by foreground and
background material.  Previously, \cite{iss90} and \cite{adl92} have made a 
related point that inferred broad linewidths of apparently quite
massive GMCs may arise due to overlapping in velocity space of 
narrower velocity distributions from individual 
smaller clouds superimposed along the line of sight.
Relatively steep increases of linewidth with size, as reported by 
\cite{lar81} and 
subsequent authors, may be obtained in observations provided that a
structure is distinguishable from its surrounding by a sufficient density
or chemical contrast;  these steeper laws correspond to what we measure with 
our 3D ROC procedure (solid lines in \ref{fig-lws}a, \ref{fig-lws2}a).

In Fig. (\ref{fig-lws}b), we show the distribution of mass with
projected size for the ROCs;  the mean logarithmic slope is nearly equal
to two -- rather than three, as would be the case for compact objects
with three comparable dimensions.  The mean slope is close to two simply
because each ROC samples along the entire line-of-sight so that mass is
nearly proportional to projected area;  note, however, that at small
scales, the masses can lie considerably above the mean fit.

It is interesting to compare the virial parameter $\alpha$ vs. $M$
distribution shown in Figure (\ref{fig-lws}) with the analogous plot
presented by \cite{ber92} analyzing the properties of apparent clumps in
four different observed clouds.  For the data sets considered in that
work, linear fits to the $\log\alpha$ vs. $\log M$ relation gave slopes
between $-0.5$ and $-0.68$.  For the model data shown in Fig.
(\ref{fig-lws}) (and for our other snapshots as well), the mean fit has
a somewhat shallower slope.  But the upper envelope of this distribution
(and those for other snapshots) has slope $\sim -0.5$ to $-0.6$.  We can
understand this upper envelope as follows:  First, the largest velocity
dispersions at a given projected scale (cf. Fig.  \ref{fig-lws}a) are
nearly independent of scale (typical logarithmic slope is $\sim 0$ to
$-0.1$).  With this, together with the mass scaling nearly as $s^2$, the
result is an upper envelope of $\alpha\propto M^{-0.5} - M^{-0.6}$.  The
relatively flat lower envelope of the $\alpha$ vs.  $M$ distribution can
be explained by the projected ROCs that sample the $\sigma_v\propto
s^{0.5}$ lower-envelope of the linewidth-size distribution (following
the true 3D linewidth-size relation), together with the approximate
$M\propto s^2$ scaling.  

All of the other model snapshots show qualitatively similar
distributions of the kinetic parameters for ROCs to those shown in
Figure (\ref{fig-lws}).  For example, we show the same distributions 
obtained for a weak-magnetic-field model snapshot (D2) in Figure
(\ref{fig-lws2});  qualitatively, all of the kinetic scalings are quite 
comparable to those obtained for the strong-magnetic-field model.
In general, for the model snapshots in Table
\ref{tbl-2}, the projections parallel to the magnetic field axes yield
slightly stronger increase of linewidth with size than do the other
projections.  For projections perpendicular to the mean field, the
ranges in the fits for the different snapshots are
$d\log(\sigma_v)/d\log(s)=0.07-0.12$,
$d\log(\sigma_v)/d\log(M)=0.03-0.08$, and $d\log(\alpha)/d\log(s)=-0.45
$ to $-0.34$ (using the same minimum surface density contrast factor
$f_c=1$ and $\rho_{min}=3\bar\rho$).  For projections parallel to the mean
field, the respective ranges for these fits are $0.11-0.19$,
$0.06-0.12$, and $-0.40$ to $-0.29$.  The fits to $d\log(M)/d\log(s)$
have a very small range, $1.83-1.89$, for all projections (using $f_c=1$
and $\rho_{min}=3\bar\rho$).

The results depend weakly on the definition of a ROC, and in
particular on $\rho_{min}$.  Reducing $\rho_{min}$ tends to yield
flatter slopes for $d\log(\sigma_v)/d\log(s)$ and
$d\log(\sigma_v)/d\log(M)$ (because velocity is anticorrelated with
density, so that additional low-density material along the line of
sight increases the dispersion closer to the maximum), and steeper
slopes for $d\log(M)/d\log(s)$ (approaching 2, the limiting form for
uniform column density), and for $d\log(\alpha)/d\log(s)$ (approaching
$-0.5$, the limiting form for velocity dispersion independent of size
and uniform column density).  Increasing $\rho_{min}$ has the opposite
effect.  The changes in slopes come about mainly from variations in the
loci of the lower envelopes of the distributions when $\rho_{min}$ 
varies;  the upper envelopes change very little, since they reflect 
the kinetic properties of ROCs which sample through the largest 
possible portion of the model cloud.

Because the projected ROC identification algorithm does not take into
account any line-of-sight information for the material in any projected
region, it should not be surprising that the velocity dispersions for
projected regions can be much larger than the velocity dispersions for
3D cubes with the same projected size.  One might argue that foreground
and background material extraneous to a principal condensation could
easily be removed based on velocity information, so that structures
identified as contrasting regions in observed molecular line $l-b-v$
data cubes would truly represent spatially coherent structures.
Examination of the line-of-sight velocity and line-of-sight position
distributions for individual projected ROCs, however, suggests that it
may in fact be difficult to eliminate foreground/background
contamination.  

To illustrate the problem, Figure (\ref{fig-line8}) shows histograms
of line-of-sight velocity (equivalent to a line profile for an
optically-thin tracer uniformly excited above $\rho_{min}=3\bar\rho $) for
regions of projected linear scale $s=L/8$.  Although some of the line
shapes are irregular, none of those meeting the ROC criterion (in this 
example) are clearly multicomponent distributions.  For comparison, in Figure
(\ref{fig-pos8}), we show the distribution of mass with position along
the line of sight.  Evidently almost every region - both ROCs and
non-ROCs -- has multiple spatial components along the line-of-sight.
Figures (\ref{fig-line16}) and (\ref{fig-pos16}) show the same
distributions for spatial regions at higher resolution; again, almost
all velocity profiles are single component, while spatial
distributions are multicomponent.  By dividing our data cubes in half
and computing velocity histograms separately for the ``front'' and
``back'' halves, we have checked that the ubiquity of single-component
velocity distributions is not an artifact of periodic boundary
conditions.  We have also checked that the phenomenon of multi-spatial
component/ single-velocity component ROCs is still prevalent even when
the density threshold $\rho_{min}$ is set higher; for example, Figure
\ref{fig-map_r10} a,b shows the velocity and position distributions
for the same model as before, but now with $\rho_{min}=10\bar\rho$.
A complementary phenomenon that we have also identified in several
model projections is that multiple velocity components in a given ROC may
correspond to a {\it single} extended spatial component --  as a consequence,
for example of two ``colliding'' clumps being viewed during 
a merger along the line-of-sight.  

The general lack of correspondence between structures in position and
velocity space in ISM models has previously been noted, based on
various sorts of analyses. For example, \cite{adl92} analyze the model
galactic disks generated from two-dimensional 
N-body cloud-fluid simulations, and show that apparent single ``clouds'' in
longitude-velocity space are often highly extended along the 
line-of-sight, and that what appears to be a single GMC in a spatial plot may 
be assigned to multiple ``clouds'' in longitude-velocity space.
\cite{pic00} show that the morphology of structures in 
position-position-velocity space (equivalent to channel maps) 
in their 3D MHD simulations is more strongly
correlated with velocity structures in physical space than with
density structures in physical space.  

The current analyses and previous work on this question do not treat 
molecular excitation and radiative transfer in detail.  Studies that
do include these complex effects will be required to reach definitive 
conclusions on the relation between maps of molecular lines and 
3D physical density-temperature-velocity cubes.
If spatially compact regions have substantially 
higher molecular excitation than more diffuse surroundings 
due to line trapping, then it is still
possible that velocity information could be used to separate
spatially-connected clumps from foreground and background
material.  Large amplitude rotation of clouds, if present, would also 
help to differentiate superposed line-of-sight clumps in the velocity domain.
Potentially, methods that use specific information about 
spectral line shapes (e.g. \cite{ros99})
may also be adapted to discriminate spatially-separated regions.
The present simplified analysis suggests, however, that
foreground and background material may at least significantly increase
the dispersion in the linewidth-size distributions for clumps identified
from molecular emission data cubes (e.g. \cite{wil94}, \cite{stu90}).

\section{Magnetic field distributions and simulated polarization}

We now turn to magnetic field structure, and begin by considering how
the distribution of the magnetic field varies for models with different
mean magnetization.  As seen in Figure \ref{fig-enev}d, the rms magnetic
field strength initially increases, due to the generation of perturbed
field by velocity shear and compression.  The distribution of the
individual components of $\bf B$ for matched Mach number model snapshots
B2,C2,D2 is shown in Figure \ref{fig-bhist}.  The
dimensionless field strength ${\cal B}\equiv B/(4\pi\bar\rho
c_s^2)^{1/2}$ that we report can be converted to a physical value using
\begin{equation}\label{eq-bconvert} 
B=1.4\times {\cal B}\times(T/10{\rm K})^{1/2} 
(n_{H_2}/10^2\cm^{-3})^{1/2} \ \mu {\rm G} .
\end{equation}
As illustrated by the figure, the component distributions are more
nearly Gaussian for the case of stronger magnetic fields; this is true
for all of the model snapshots, although the distributions in the
high-$\beta$ (low-$B_0$) cases do become more Gaussian in time.  For the
weak-field models, the dispersion in each component of the magnetic
field is larger than the mean field component.

Because magnetic fields are measured via the Zeeman effect with
different atomic and molecular tracers in different density regimes,
it is interesting to analyze how the mean field strength in
simulations may depend on density.  Since the magnetic field is weaker
and less able to resist being pushed around by the matter in
the $\beta=0.1,1$ (C and D) simulations, one expects that the field
strength will have stronger density dependence for these models than
for the $\beta=0.01$ simulation.  This is indeed the case, as can be seen 
in Figure \ref{fig-bvsrhoA}.  Particularly at densities below the mean, the 
magnetic field strengths in the high-$\beta$ models are strongly density
dependent; the low-density slope of $d\log B/d\log \rho$ for these 
models is near the value $2/3$ associated with a constant ratio of 
mass to magnetic flux and isotropic volume changes.  

At high densities (above $\sim 10\bar\rho$) the relatively flat slope of
the $\beta=0.01$ model increases, becoming comparable to the slopes of
the $\beta=0.1,1$ models.  Figure \ref{fig-bvrhigh} shows the
high-density $B$ vs. $\rho$  dependence, for various model snapshots;
fits  for fiducial density $n_{H_2}$ in the range $10^3-10^4 \cm^{-3}$
(i.e. $\rho/\bar\rho=10$ to $100$) yield slopes $0.3-0.7$ for $d\log
B/d\log \rho$.  Only the Mach-9 $\beta=1$ model yields a
high-density-regime slope as steep as the isotropic contraction limit.
The other snapshots have slopes 0.3-0.5, which may be compared with the
slope 0.47 found from a compilation of Zeeman measurements at high
densities $n_{H_2}=10^2-10^7\cm^{-3}$ \citep{cru99}.  The values of the
mean $B^2$ in any density regime generally increase with increasing mean
net magnetic flux  $B_0$ (i.e. decreasing $\beta$), but because there is
significant dispersion about the mean $B^2$, there is considerable
overlap of the $1-\sigma$ deviation regions among the different model
snapshots (Fig. \ref{fig-bvsrhoA}).

The numerical results on the $B$ vs. $\rho$ relation presented by 
\cite{pad99} (see their Fig. 7) are qualitatively similar to our 
results, with some differences apparent at the high density end.  
The lower Mach number in their low-$\beta$ model
compared to their  high-$\beta$ model likely accounts for its relatively
weaker increase of $B$ with $\rho$ at high density, compared to our 
results.  We also differ
with those authors regarding the astronomical implications of the
numerical results.  In particular, we do not attempt a comparison of 
the low-density end of the $B$ vs. $\rho$ distributions with observations
made in the diffuse ISM, because (a) the physical regime modeled by 
the simulations is not appropriate for the diffuse ISM (where thermal 
pressure is comparable to, rather than much smaller than, 
$\langle \rho v^2\rangle$);
and (b) the transformation from simulation to physical variables for 
local magnetic field values involves multiplying by the mean magnetic 
field $B_0$ on the largest scale, and this need not be the same in the 
diffuse and cold ISM ($\beta$ parameterizes this mean field strength).  
We conclude that the $B$ vs. $\rho$ relations obtained from
simulations do not at present constrain the value of $\beta$.  
At high densities,
all models (either weak or strong $B_0$ on the large scale) yield
slopes which are consistent with high-density molecular Zeeman
observations.  At very low densities, where the predictions of models
with varying $\beta$ {\it do} differ, estimating $B$ {\it within}
clouds would be difficult, since HI Zeeman observations probe the high
columns of foreground and background material, rather than the low
column of cloud material (although velocity information may help; cf.
\cite{goo94}). The field strengths in the low density regions {\it within } 
molecular clouds may in fact be systematically higher than those at 
comparable density in the diffuse ISM.

For all the snapshots, there is significant dispersion in the total
magnetic field strength.  In addition to this overall dispersion in
magnitude, there is a dispersion in the magnetic field vector
direction which increases with decreasing strength of the mean field
component $B_0$, simply because fixed amplitude fluctuations have
larger relative amplitudes compared to a weak mean magnetic field.
The dispersion in field directions has important consequences for any
observational measurement of the mean magnetic field via Zeeman
splitting.  Observations of Zeeman splitting at any position on a map
yield the line-of-sight average value for the line-of-sight magnetic
field, weighted at each point along the line-of-sight by the local
excitation.  When a given line of sight has many fluctuations in the
{\it direction} of the magnetic field, the average value of $\langle
B_{los}\rangle$ will be {\it small}, even if individual local
components of the field have large magnitudes.

To demonstrate how the averaged line-of-sight field components vary
with mean field strength and observer orientation, we depict in Figure
\ref{fig-bzmap} an overlay of $\langle B_{los}\rangle $ on the column
density for three model snapshots with matched Mach number.  In Figure
\ref{fig-bzvscol}, we plot the values of $\langle B_{los}\rangle $ vs.
column density of dense gas.  The Figures show, unsurprisingly, that
the line-of-sight-averaged magnetic field strengths are greatest when
the mean field $B_0$ is largest and is oriented along the observer's
line-of-sight (top left panel).  For the weaker-field models, the
average line-of-sight field is lower, and there is larger dispersion.
For all the snapshots, there is considerable dispersion in the values
of $\langle B_{los}\rangle $ on the map, and the largest values do {\it not}
correspond to the positions of highest column density; in fact, there
is some tendency of line-of-sight-averaged field to {\it
  anticorrelate} with column density.  Thus, although the local field
strength $|B|$ increases with density (cf. Figs \ref{fig-bvsrhoA},
\ref{fig-bvrhigh}) and may be much larger than the volume-averaged
mean field $B_0$ for the entire box, line-of-sight superpositions of
non-aligned vector components produce average line-of-sight field
strengths closer to the large-scale volume-averaged value.

It is well known that it is difficult to detect the Zeeman effect in 
molecular clouds (e.g.
\cite{hei93}) because the frequency splitting is small when the field is
weak.  This, coupled with the possibility (cf.  Fig.  \ref{fig-bzvscol})
that an impractically large number of measurements might be required to
obtain statististically-significant results for the large-scale field,
underscores the importance of supplementing programs of direct detection
with other methods for estimating the mean field strength.  
Long before direct Zeeman detections were first made, \cite{cha53}
estimated mean spiral-arm field strengths $B_0$ from the mean gas
density, line-of-sight velocity dispersion, and the dispersion in
orientations of the magnetic field in the plane of the sky.  The field
line orientation is taken to be traced by the polarization direction for
background stars, which occurs provided that the dust grains producing
the intervening extinction are aligned with short axes preferentially
parallel to $\bf B$, and so preferentially extinguish linear
polarizations perpendicular to $\bf B$.

The \cite{cha53} (hereafter CF) estimate is based on the fact that for
linear-amplitude transverse MHD (Alfv\'en) waves, $B_p =
\sqrt{4\pi\bar\rho} |\delta {\bf v}|/(|\delta {\bf B}|/B_p).$ Here $B_p$
is the projection of the mean magnetic field on the plane of the sky, and
$\delta{\bf  B}$ and $\delta {\bf v}$ are the components of the magnetic
and velocity perturbations in the plane of the sky transverse to ${\bf
B}_p$.  If the interstellar polarization is parallel to
the local direction of ${\bf B}_p$, then the ratio $|\delta {\bf B}|/B_p$
in the denominator may be replaced by the dispersion $\delta \phi$ in
polarization angles (for small angle/low amplitude perturbations).  With
the further assumption that the true velocity perturbations are isotropic,
then the dispersion in the transverse velocity $|\delta {\bf v}|$ is
equal to the rms line-of-sight velocity $\delta {v_{los}}$.
We thus obtain 
\begin{equation}\label{QDEF}
B_p = {\cal Q} \sqrt{4\pi\bar\rho}\ \delta {v_{los}}\ \delta\phi^{-1}
\end{equation}
where, for CF's field model, ${\cal Q} = 1$.  Modifications to the CF
formula allowing for inhomogeneity and line-of-sight averaging are
discussed by \cite{zwe90} and \cite{mye91}, respectively.  Both of these
effects (and others;  see \cite{zwe96}) tend to reduce ${\cal Q}$.

Potentially, the CF ``polarization-dispersion'' method can be used to
estimate plane-of-sky magnetic field strengths on scales within
turbulent interstellar clouds.  It may also be possible to combine these
results with Zeeman measurements to estimate the total magnetic field
strength \citep{mye91,goo94}.  In order to evaluate the ability of the
CF method to measure mean plane-of sky field strengths, we provide a
first (simplified) test of it using our model turbulent clouds.  
For this test, we have created simulated polarization maps for
each cloud by integrating the Stokes parameters along the line-of-sight
over a projected grid of positions, assuming the polarizability in each
volume element is proportional to the local density.  The details of
this procedure, together with a more extensive discussion of simulated
polarization distributions, will appear in a separate publication
\citep{ost00}.  

For two projected model snapshots (B2 with $\beta=0.01$ and D2 with
$\beta=1$ projected along $\hat z$), Figures \ref{fig-imagelb} and
\ref{fig-imageld} show examples of the polarization maps overlaid on
color scale column density maps.  The analogous map (not shown) for the
model C2 ($\beta=0.1$) looks quite similar to Figure \ref{fig-imageld}.
From the figures, it it immediately clear that the model with a stronger
mean magnetic field $B_0$ has more ordered polarization directions and
larger typical values of the fractional polarization, compared to the
model with a weaker mean magnetic field.  These trends are as expected:
a weaker mean field has lower tensile strength, so that for a given
level of kinetic energy the Reynolds stresses will produce larger
fractional perturbations in the magnetic field -- corresponding to
larger fluctuations in projected line-of-sight averaged position angle.
Also, because of the larger dispersion in local polarization directions
along any line of sight, cases with weaker mean magnetic fields will
show lower net polarization through the cloud (from the line-of-sight
averaging of the varying local vector directions).  While local 
(line-of-sight averaged) polarization
directions may have any orientation with respect to local projected
surface density, there is some tendency for the large-scale 
projected density and large-scale polarization directions to align in
the high-$\beta$ (but not low $\beta$) models, because the magnetic field 
and density are both strongly sheared and compressed by the large-scale,
large-amplitude velocity field.

In Figure \ref{fig-polang}, we show the distributions of polarization
angle at various ``observer'' orientations 
for models with matched kinetic energy and mean magnetic field
strengths at three different levels ($\beta=0.01,0.1,1$, corresponding
to fiducial $B=14, 4.4,$ and $1.4\mu\G$ from eq.  \ref{eq-bconvert}).
As is clear from the Figure, only the strong-field model has
significantly correlated directions in the simulated polarization
vectors.  This is expected, since only this model has
perturbed magnetic energy lower than the mean magnetic energy; the
ratios are $\delta B^2/B_0^2$ are 0.27, 4.0, and 12, respectively, for
the snapshots presented.  

For the cases shown in Figure \ref{fig-polang} where the angle
dispersion $\delta \phi$ is $25^\circ$ or less (i.e. the $\beta=0.01$
projections at $i=0, 30, 45,$ and $60^\circ$), we have compared the
known value of the mean plane-of-sky magnetic field 
$B_p\equiv B_0\cos(i)$ with the
Chandrasekhar-Fermi estimate.  We find that $\cal Q$ (see equation
(\ref{QDEF})) is in the range 0.46-0.51.  This suggests that the CF
estimate, modified by a multiplicative factor $\sim 0.5$ 
to account for a more complex magnetic field and density
structure, can indeed provide an accurate measurement of the
plane-of-sky magnetic field when the polarization angle fluctuations
are relatively small.  The method fails, however, when the
polarization angle fluctuations are large.
We will present a more comprehensive analysis of this promising
diagnostic in a separate publication.

\section{Summary and discussion of structural analyses}

With modern high-performance computational tools, it is possible to
create and evolve simulated dynamical representations of turbulent, magnetized
clouds at comparable plane-of-sky spatial resolution to that of
radio-wavelength observational maps of GMCs.  This paper reports on the
properties of a set of such simulations. 

We start by briefly summarizing (\S3) the results on energy evolution
in our simulations.  We confirm the conclusions from our previous work
that turbulent decay is rapid even in magnetized models, finding that
an interval of only 0.4-0.8 flow crossing times is sufficient to
reduce the total turbulent energy by a factor two from its initial
value; the corresponding physical time for GMC parameters is only a
2-4 million years.  We also confirm that in situations
where turbulence is not replenished, the criterion for a cloud to collapse 
gravitationally depends only on whether it is sub- or super-critical
with respect to its mean magnetic field;  the characteristic 
collapse time in the latter case is $\sim 6$ Myr for GMC parameters. 

Following the presentation of energetics, the bulk of the
paper (\S\S 4-6) is concerned with developing tools for structural 
analyses and applying them to our simulated data cubes.
Although simplified in their
treatment of small scales (ambipolar diffusion is neglected) and thermal
properties (a constant gas temperature is assumed), the 3D data cube
``snapshots'' from our numerical experiments provide a detailed 
portrait of the
density, velocity, and magnetic field structure in the simulated clouds.
This structural portrait is dynamically self-consistent in that it is an
instantaneous solution to the full time-dependent MHD equations: the
density and magnetic field variables have evolved in response to a
(time-dependent) turbulent velocity field, which itself has evolved
subject to gas pressure gradient forces, magnetic stresses, and
self-gravity.

Model cloud snapshots from simulations provide a unique opportunity to
(i) explore the intrinsic
character of 3D structure in magnetized gaseous systems subject to
supersonic turbulence, and (ii) determine which aspects of the observed
properties of GMCs (from 2D plane-of-sky integrated maps or $l-b-v$ data
cubes) can be explained as a manifestation of their internal turbulence.
The possibilities for such exploration are enormous; for   
practical purposes, we have limited the scope of this paper to 
three groups of
analyses.  We consider: 
(1) the distributions of mass, volume, and
area as functions of volume density and column density (\S 4); (2)
the distributions of 
velocity dispersion, mass, and virial parameter $\alpha$ as a
function of the spatial scale for zones in projected maps and cells in
3D cubes (\S 5); (3) the distributions of magnetic field strength vs.
local volume density, line-of-sight-averaged line-of-sight magnetic
field vs. column density, and distribution of simulated polarization
angles (\S 6).  For each of these analyses, we compare sets of cloud
snapshots in which the turbulent Mach number is matched, and the large
scale mean magnetic field strength $B_0$ varies by a factor ten, also
allowing for different ``observer'' viewing angles.  The rms tubulent
velocities for the model snapshots are $\sigma_v=1-2\ \kms$, and the
mean magnetic field strengths are $B_0=1.4-14\ \mu\G$, assuming fiducial
GMC parameters for volume-averaged density $n_{H_2}=100\ \cm^{-3}$ and
temperature $T=10\K$.

The main results of these structural analyses are as follows:

\par
\begin{figure}
\epsscale{1.}
\plotone{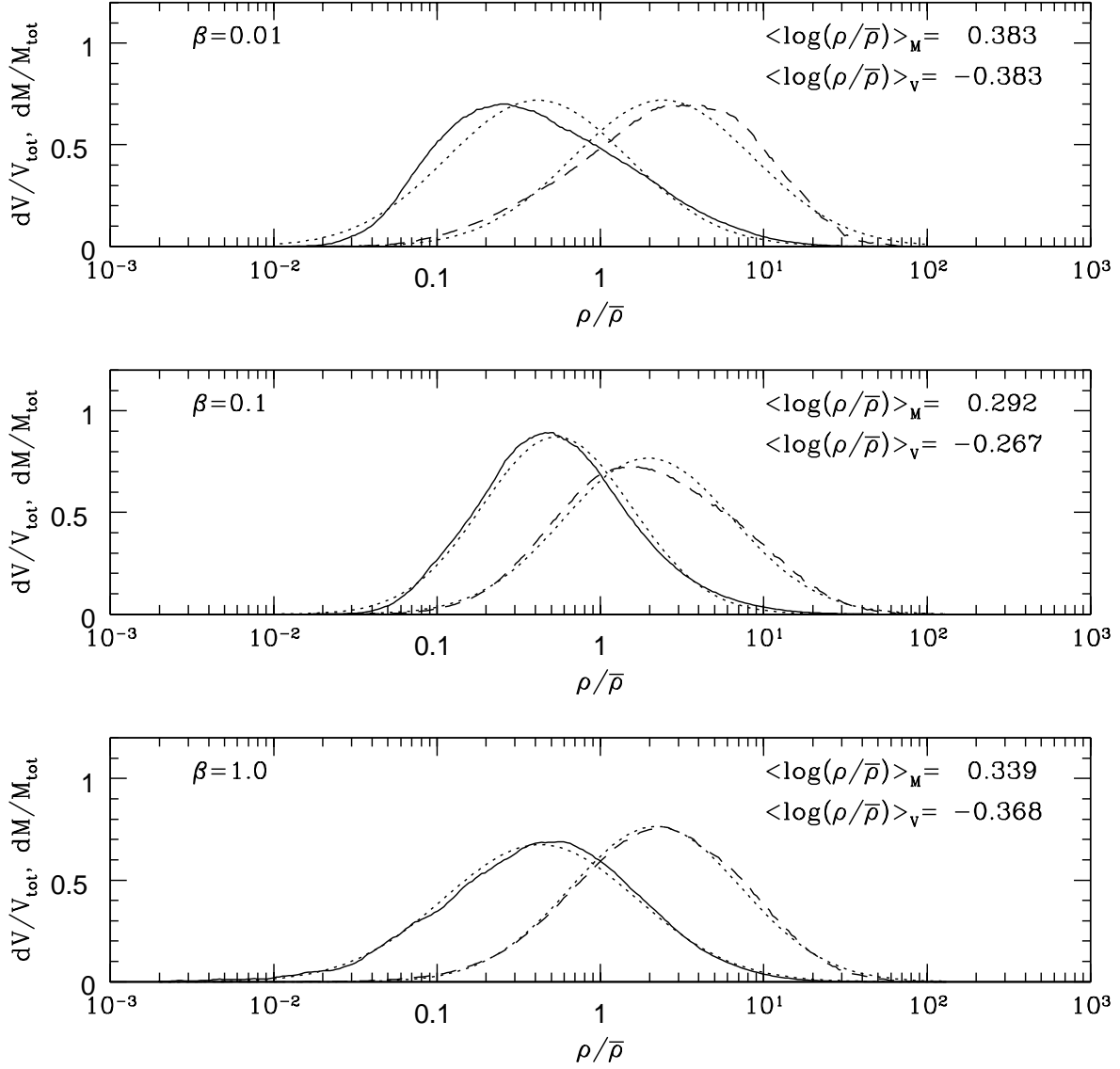}
\caption{Comparative statistics of volume density in three model snapshots
(B2, C2, D2 from Table 2) with
matched Mach numbers ${\cal M}\sim 7$. 
Solid curves show {\it fraction of volume} as a function of
density relative to the mean ($\rho/\bar\rho$); 
dashed curves show {\it fraction of mass} as a 
function of $\rho/\bar\rho$.  Dotted curves show lognormal distributions 
with the same mean and dispersion as in each model snapshot. 
\label{fig-rhohist}}
\end{figure}

\begin{figure}
\epsscale{.5}
\plotone{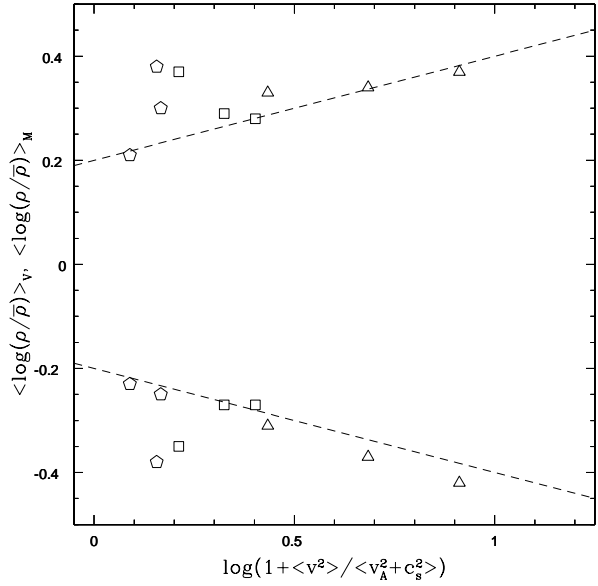}
\caption{Mass- and volume- weighted mean values of the logarithmic density
contrast as a function of mean square fast magnetosonic 
Mach number from model snapshots in Table \ref{tbl-2}.  
Pentagons, squares, and triangles
correspond to $\beta=0.01,0.1,$ and $1$, respectively.
Dashed lines show 
$\left<\log(\rho/\bar\rho)\right>_{M,V}=\pm0.2 (\log(1+ {\cal
  M}_F^2) +1)$
\label{fig-rho_vs_mach}}
\end{figure}

\noindent{\bf 1.}  The distribution of volume densities follows an
approximately log-normal form, with densities of typical mass elements 
$\langle \rho\rangle_M$ compressed by
a factor $\sim 3-6$ times the volume-averaged density $\bar\rho\equiv
M/L^3$ for our sets of snapshots with Mach number ${\cal
M}=\sigma_v/c_s$ in the range $5-9$ (see Fig. \ref{fig-rhohist} and
Table \ref{tbl-2}).  This typical density contrast is comparable to that
inferred for the concentrations in GMCs ($\langle \rho/\bar\rho\rangle_M
\sim 6-8$) from $^{13}$CO molecular-line studies (cf. Paper III;
\cite{bal87,wil95}).  The corresponding rms mass-weighted dispersion in
$\rho/\bar\rho$ is $\sim 3-13$. Although the density contrast generally
increases with the value of the fast-magnetosonic Mach number ${\cal
M}_F\equiv\sigma_v/\langle c_s^2+v_{A}^2\rangle ^{1/2}$ 
(see Fig. \ref{fig-rho_vs_mach}),
there is no obvious one-to-one functional relation between ${\cal M}$,
$\beta\equiv c_s^2/v_{A,0}^2$, and the density contrast.  In particular, the
result obtained by \cite{nor99} for purely hydrodynamic quasi-steady
turbulence of the relation between the density contrast and the Mach
number $\cal M$ does not carry over for (evolving) MHD turbulence.  When
${\cal M}\sim 5-9$, the \cite{nor99} quasi-steady
hydrodynamic-turbulence result would predict mass-weighted means and
dispersions of $\rho/\bar\rho$ in the range $\sim 7-21$ and $\sim
18-95$, respectively, larger than the range we find for our MHD models.
Further investigation would be required to determine whether, for
quasi-steady MHD turbulence, it is possible to find a clean functional
relation between the mean (and dispersion) of $\rho/\bar\rho$ and the
dimensionless parameters $\cal M$ and $\beta$ that is independent of the
particular instantaneous turbulent power spectrum. Since, however, we expect 
that cold-ISM turbulence is subject to significant transient effects, 
and in addition large ``cosmic variance'' may result from low-$k$ dominance 
of the power spectrum, a  
one-to-one relation of this kind would probably not be realized in GMCs
in any case.

\par
\begin{figure}
\epsscale{1.}
\plotone{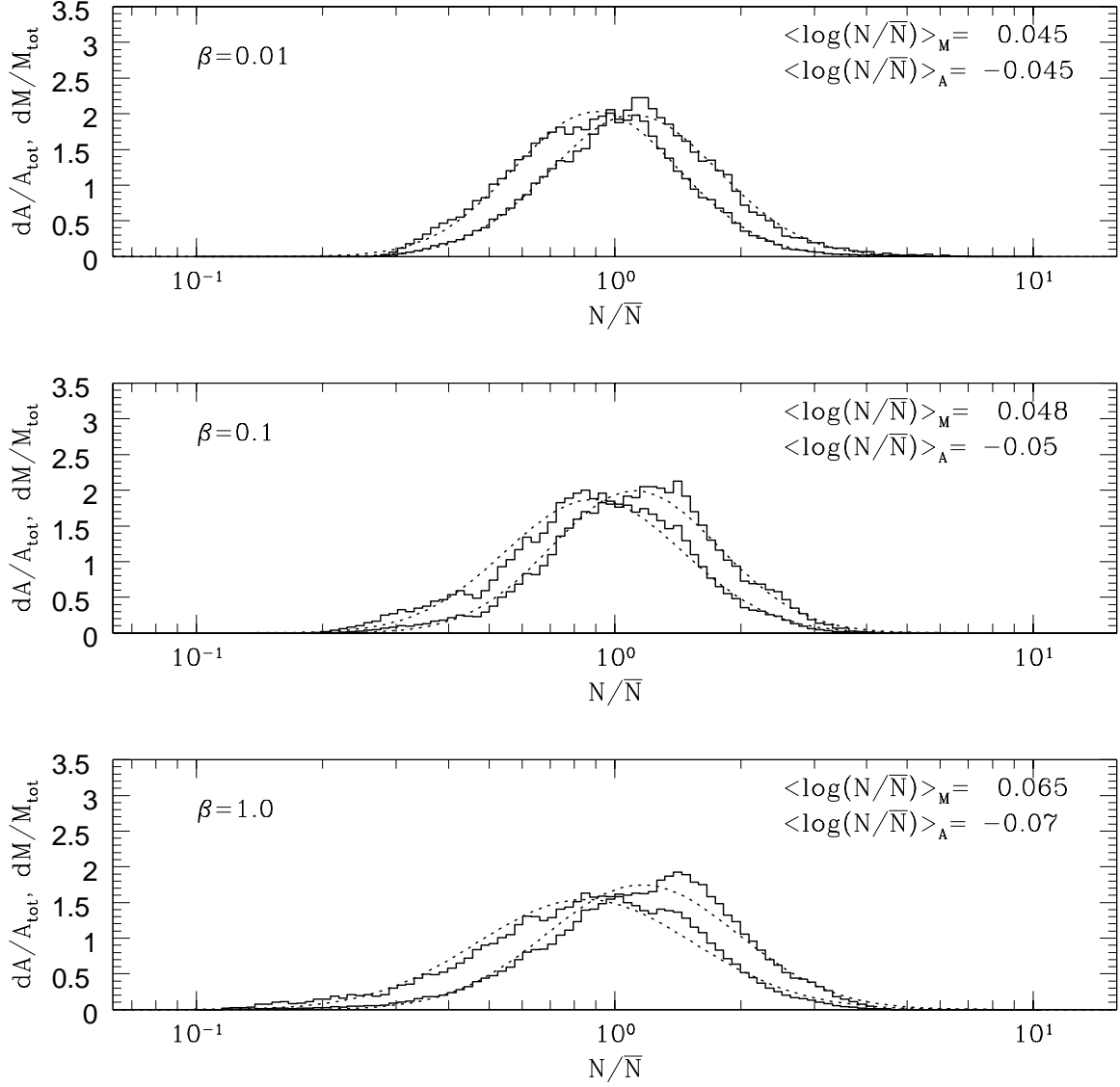}
\caption{Comparative statistics of column density in three model snapshots
(B2, C2, D2 from Table 2) with
matched Mach numbers. Projection is along the $\hat z$ axis (perpendicular to
the magnetic field).  In each frame, left-displaced 
curves show {\it fraction of projected area}
as a function of column density relative to the mean ($N/\bar\rho L\equiv
N/\bar N$); 
right-displaced  curves show {\it fraction of mass} as a 
function of $N/\bar N$.  Dotted curves show lognormal distributions 
with the same mean and dispersion as in each model snapshot. 
\label{fig-colhist}}
\end{figure}

\begin{figure}
\plotone{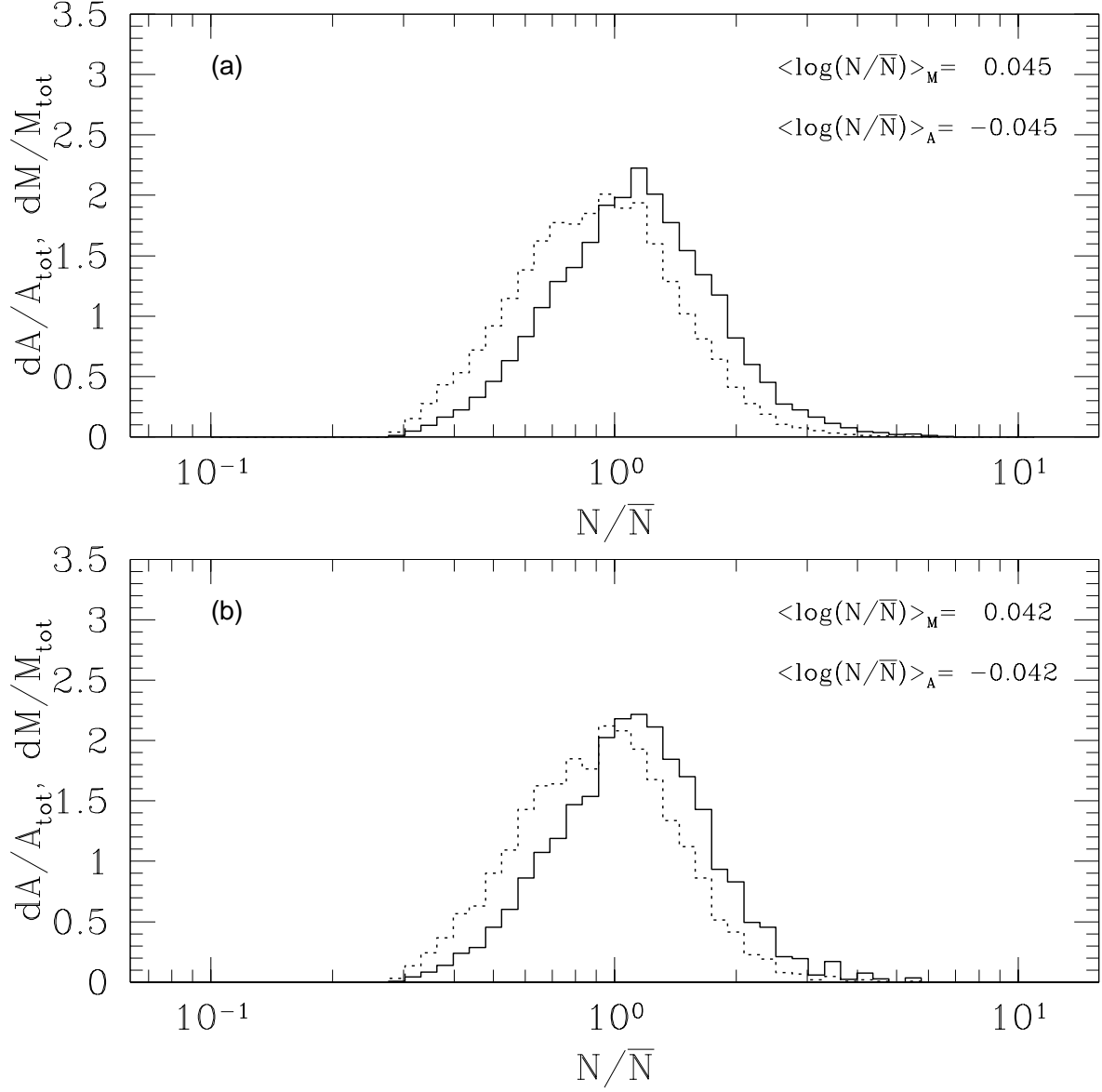}
\caption{Comparative statistics of column density at full simulation 
resolution (a) and at resolution a factor four larger in linear scale (b);
simulation data is from snapshot B2.
\label{fig-colres}}
\end{figure}

\begin{figure}
\epsscale{1.}
\plotone{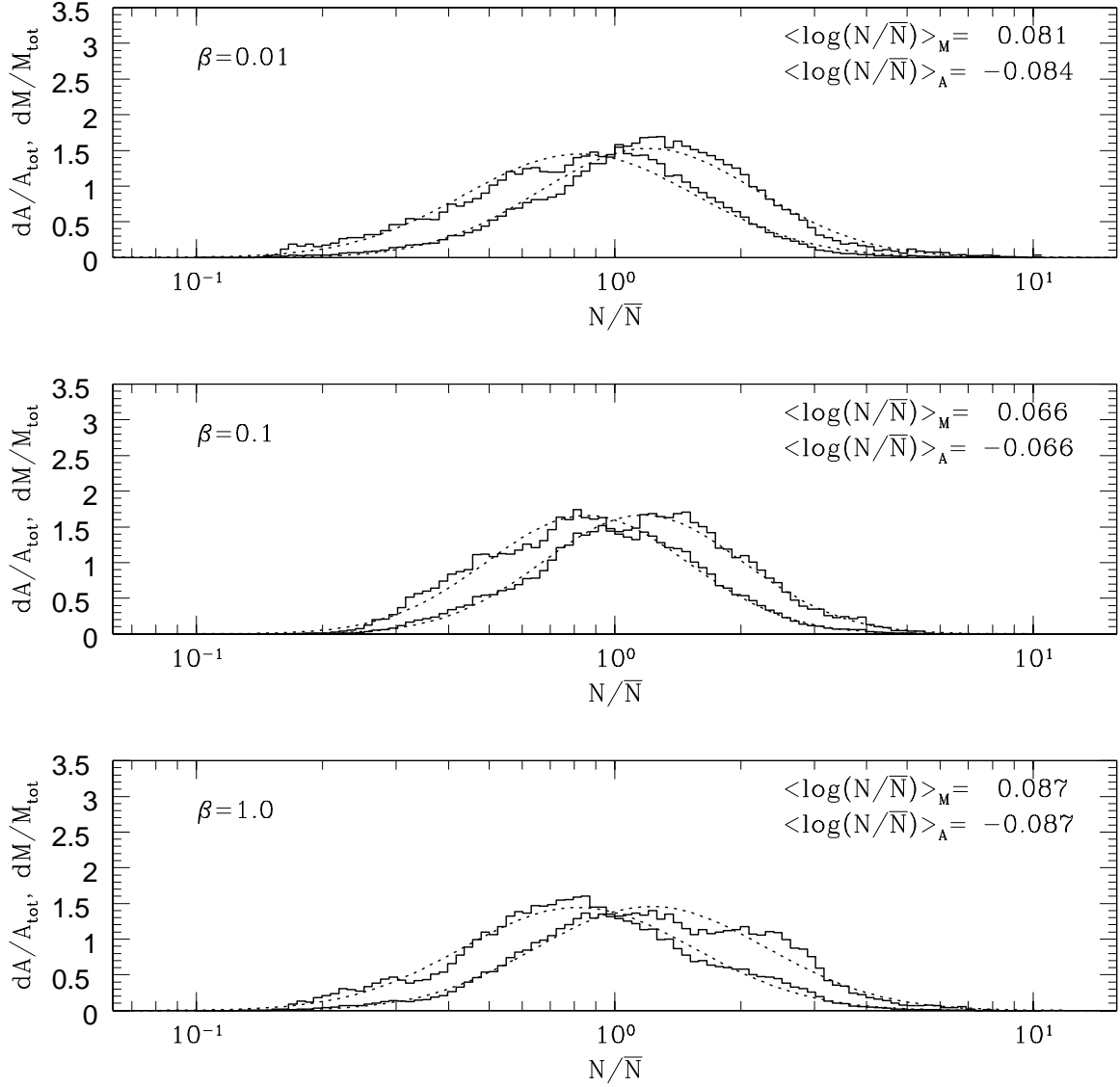}
\caption{Same as in \ref{fig-colhist}, except line-of-sight integration
is only over $z>L/2$.
\label{fig-colhistfront}}
\end{figure}

\begin{figure}
\epsscale{1.}
\plotone{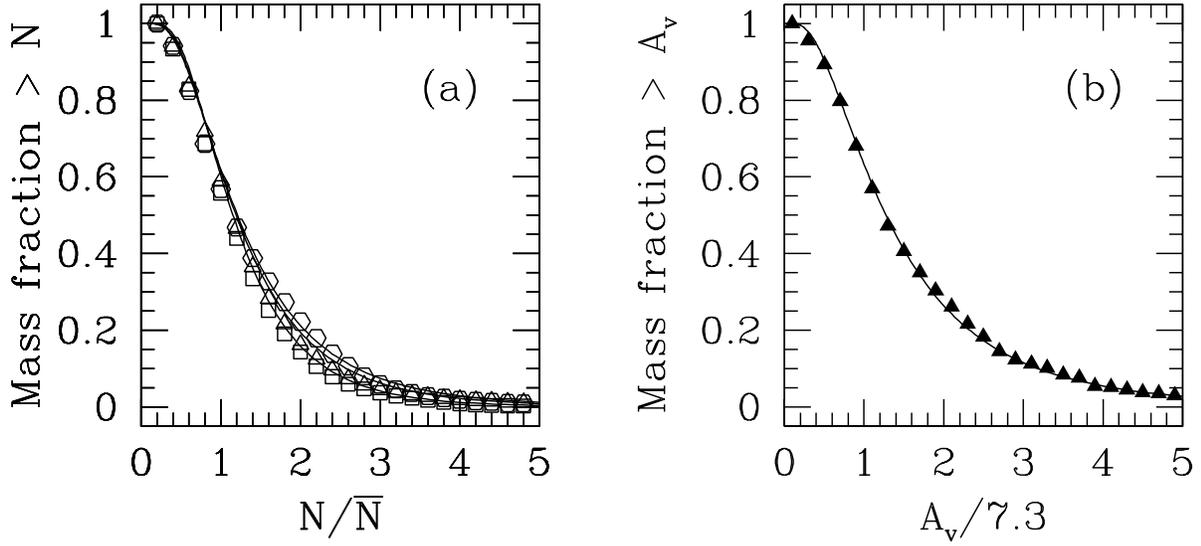}
\caption{Comparison of (a) cumulative column density distribution
  for simulated cloud model (data from Fig. \ref{fig-colhistfront},
  with hexagons, squares, and triangles marking the $\beta=1,0.1,0.01$
  model distributions) and (b) cumulative extinction distribution for
  the cloud IC5146.  Dotted curves in both panels show cumulative
  log-normal fits.
\label{fig-cumcol}}
\end{figure}

\noindent{\bf 2.} The distributions of column densities $N$ also
follow an approximately log-normal form, with mean logarithmic
contrasts $\langle \log(N/\bar N)\rangle $ an order of magnitude
smaller that the mean logarithmic density contrasts $\langle
\log(\rho/\bar\rho) \rangle$ (see Figs. \ref{fig-colhist} --
\ref{fig-cumcol} and Table \ref{tbl-2}).  The mass-weighted mean column
density is thus just 7-35\% greater than the area-weighted column
density $\bar N\equiv \bar\rho L= M/L^2$, and the mass-weighted
(area-weighted) dispersion in $N$ is 0.3-0.8 (0.3-0.6) times $\bar N$,
for models with the range of Mach numbers and magnetic field strengths
we have analyzed.  Large-scale spatial correlation of the density
perturbations is indicated by the log-normal, rather than Gaussian,
form of the column density distributions.  These large-scale spatial
density correlations are associated with large-scale correlations in
the velocity and magnetic fields (which, for the present models, are
input in the initial conditions).

\par
\begin{figure}
\plotone{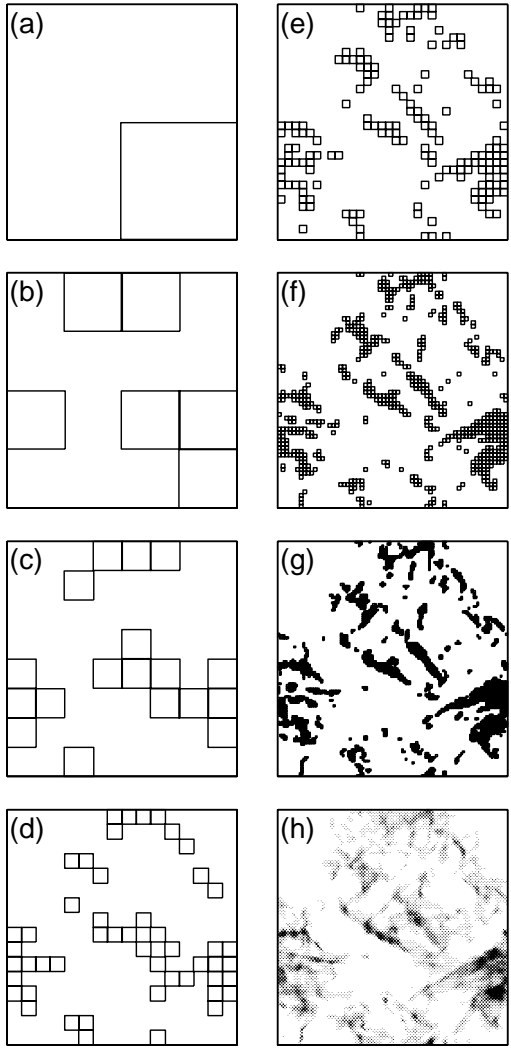}
\caption{Identification of regions of contrast (ROCs) as a function of
spatial scale for data from model snapshot B2. Panels (a)-(g) outline
regions that meet the contrast criterion at increasingly fine spatial
resolution.  A greyscale representation of the projected density is shown
in (h) for comparison.
\label{fig-roc}}
\end{figure}

\begin{figure}
\epsscale{.8}
\plotone{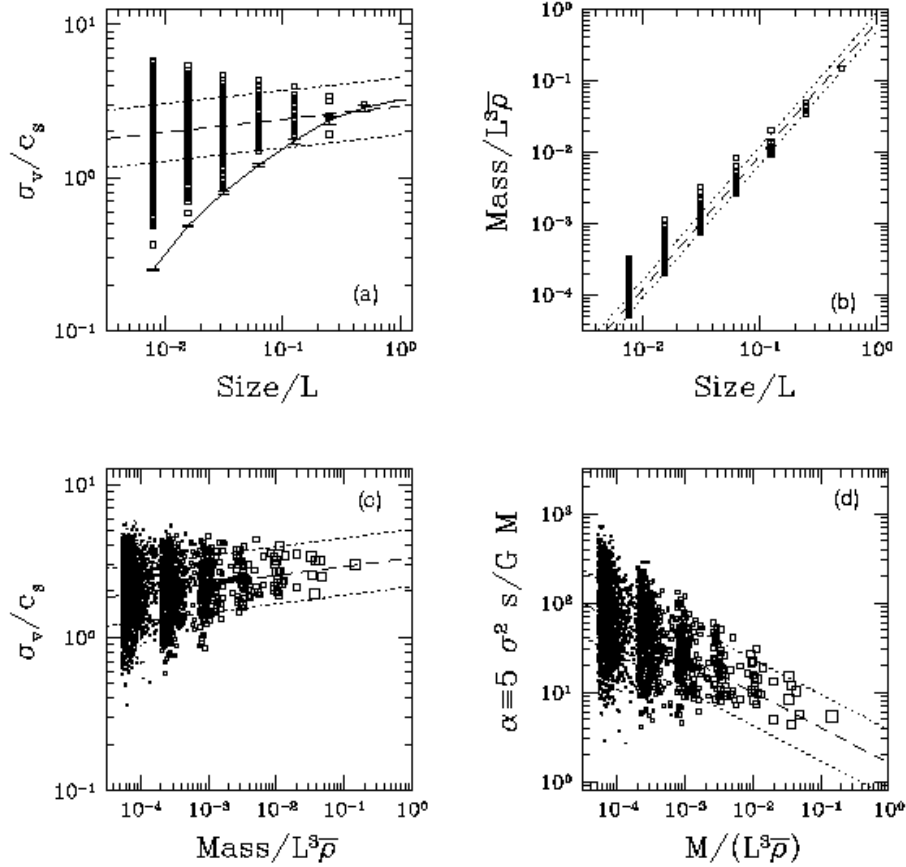}
\caption{Scale dependence of kinetic quantities in a projected map.
In (a)-(d), each point represents one of the square 
regions of contrast (ROCs) identified in Fig. (\ref{fig-roc}), with 
edge size $s$.  
In (a), we plot {\it vs.} size $s$ (normalized to the box length $L$)
the dispersion of the line-of-sight velocity $\sigma_v$ in each projected 
square ROC, and also show (solid line) the mean 
dispersion in line-of-sight velocity for
3D cubes of side $s$.  In (b), we plot {\it vs.} $s/L$  
the mass $M$ of each ROC (normalized to the total mass in the simulation
box $\bar \rho L^3$.  In (c),
we show $\sigma_v$ vs. $M$.  In (d), we show the virial parameter $\alpha$
vs. $M$.
In each frame, dashed lines represent linear least-squares fits to the 
data;  dotted lines represent 1-sigma deviations from the fit. 
In (c) and (d), we plot points from different-sized regions with different
expansion factors.
\label{fig-lws}}
\end{figure}

\begin{figure}
\plotone{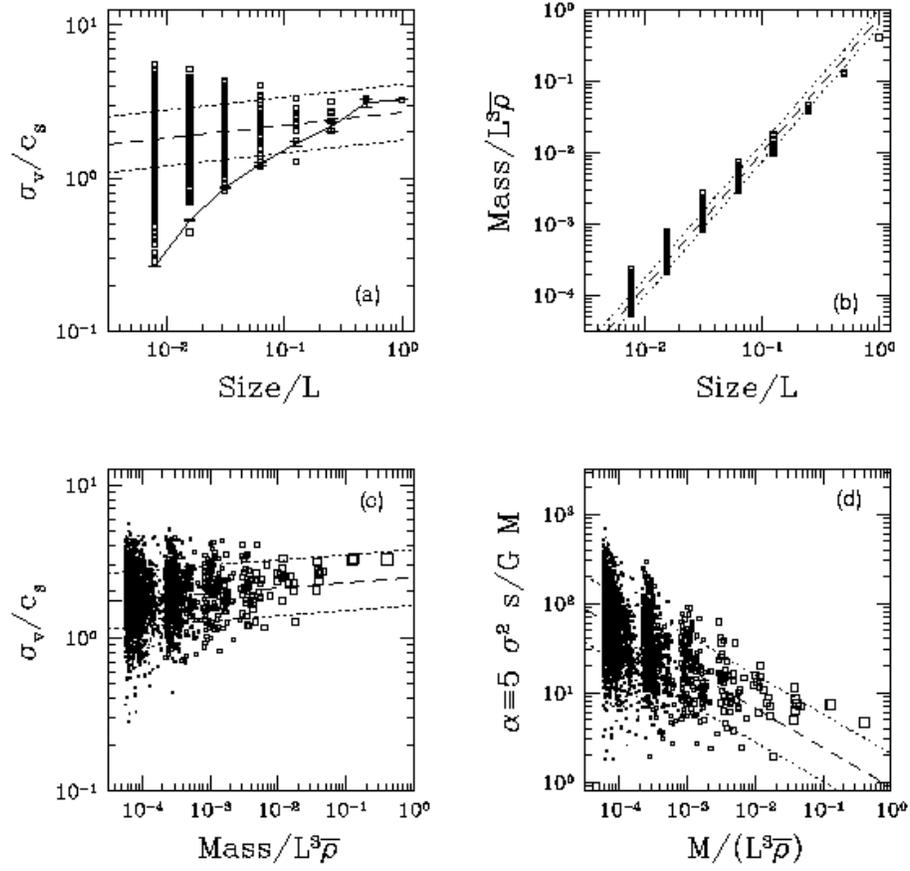}
\caption{Same as in Fig. \ref{fig-lws}, for model snapshot D2.
\label{fig-lws2}}
\end{figure}

\begin{figure}
\plotone{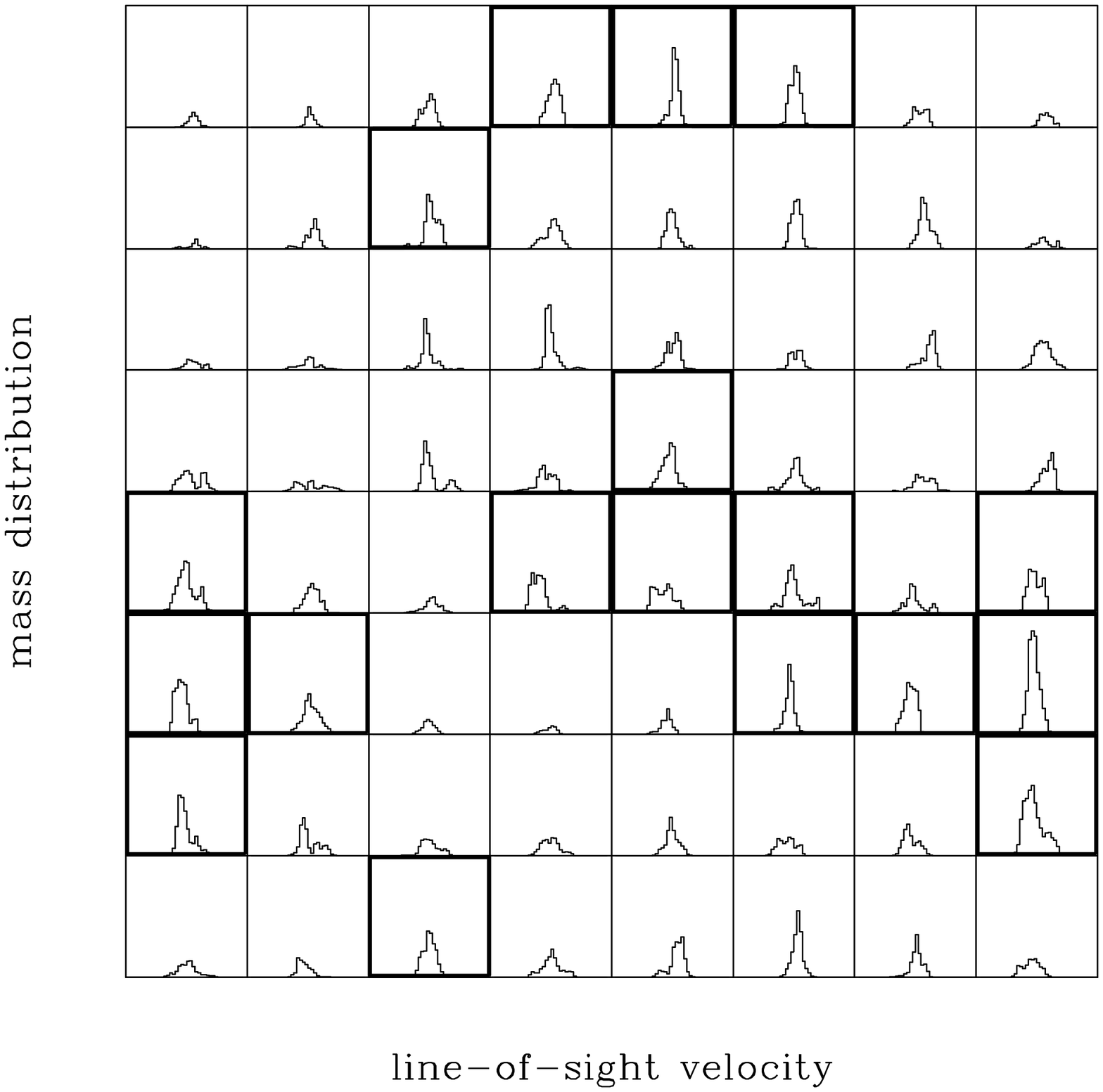}
\caption{Distribution of mass with line-of-sight velocity 
for model snapshot B2 projected 
in the $\hat z$ direction, for regions of linear size $L/8$.  
Regions meeting surface density 
contrast criterion (cf. Fig. \ref{fig-roc}) are indicated 
with heavy outlines.
\label{fig-line8}}
\end{figure}

\begin{figure}
\plotone{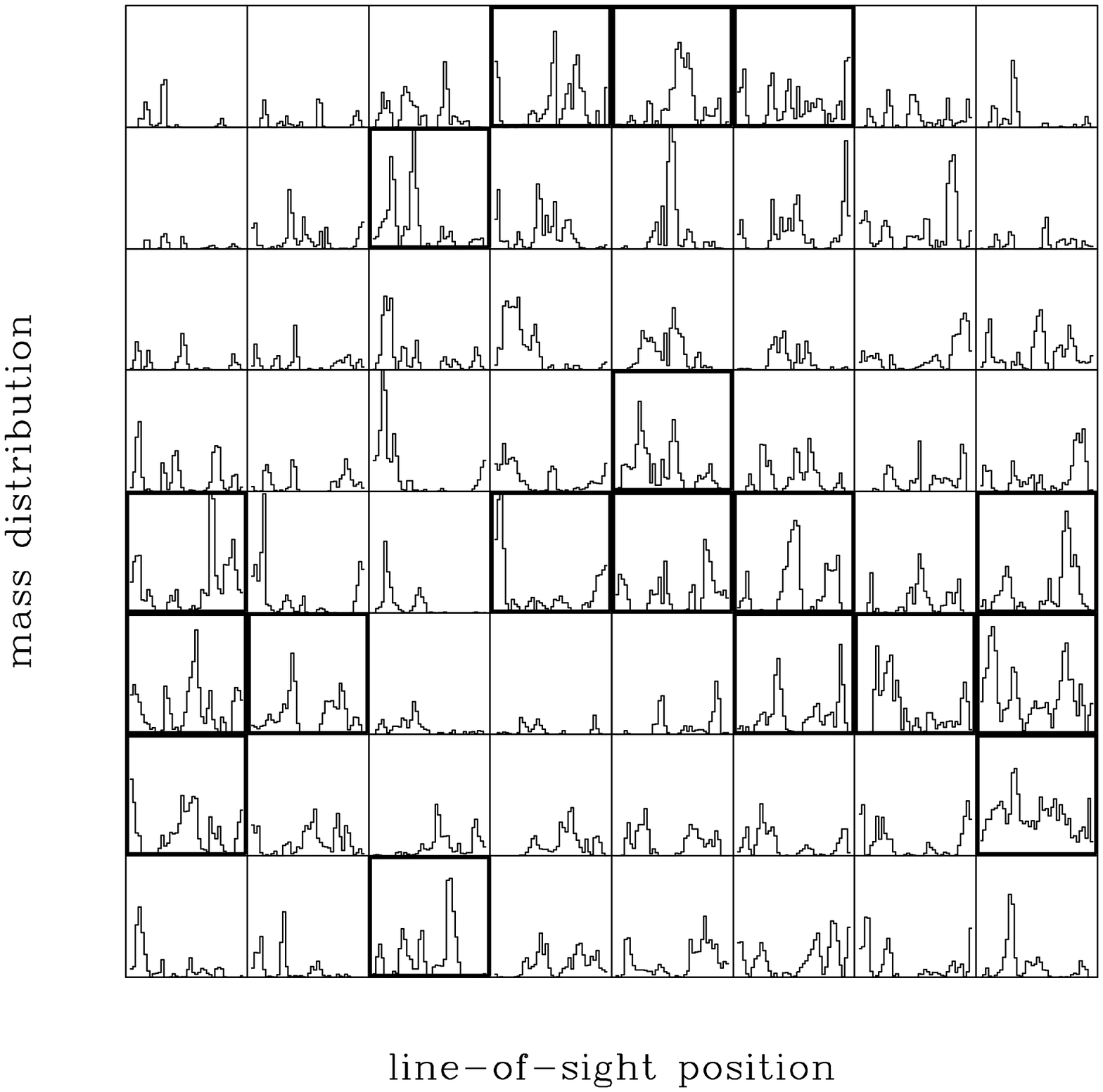}
\caption{Distribution of mass with line-of-sight position 
for model snapshot B2 projected 
in the $\hat z$ direction, for regions of linear size $L/8$.  
Regions meeting surface density 
contrast criterion (cf. Fig. \ref{fig-roc}) are indicated 
with heavy outlines.
\label{fig-pos8}}
\end{figure}

\begin{figure}
\plotone{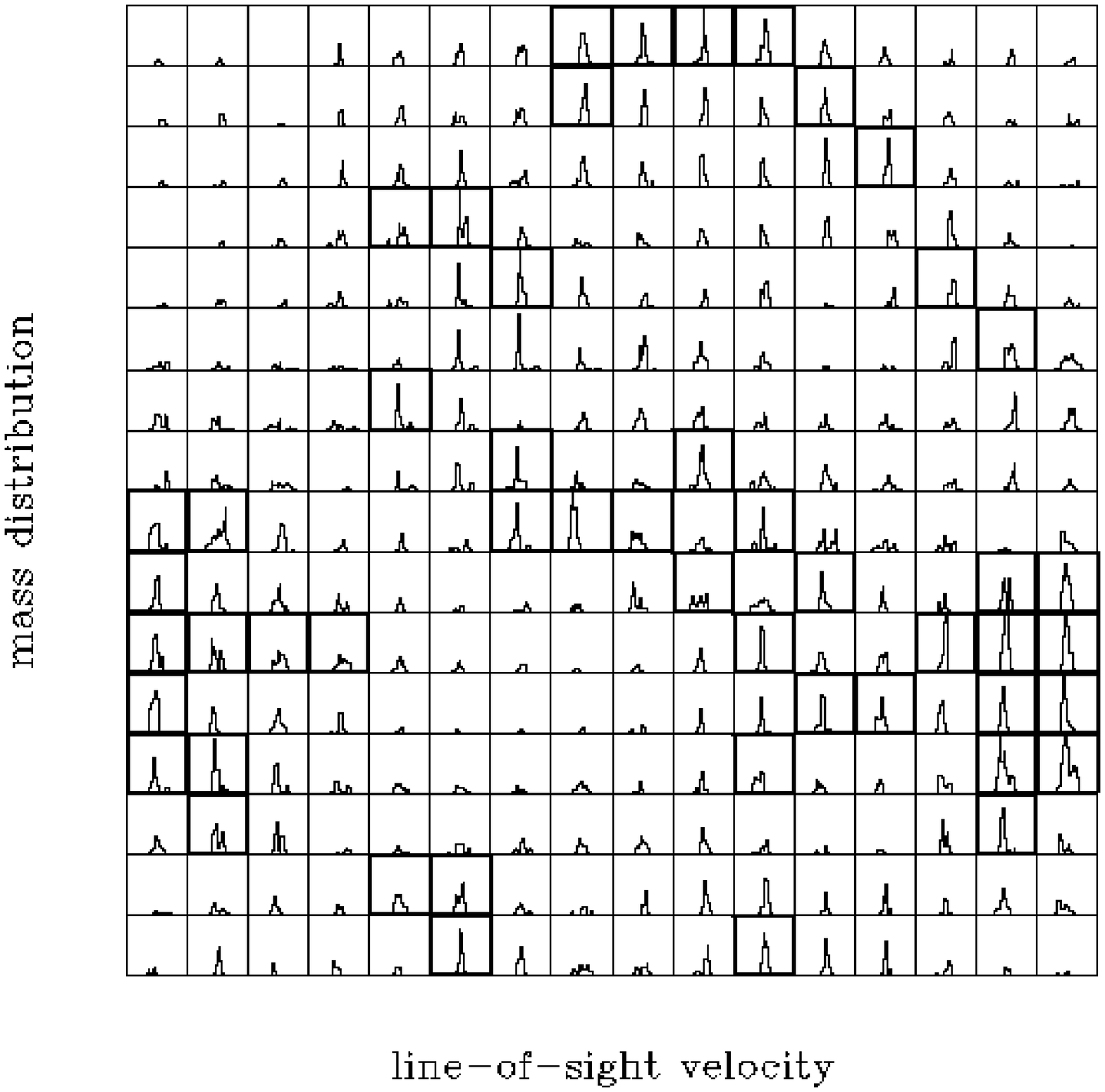}
\caption{Same as Fig. (\ref{fig-line8}), for projected region size 
$s=L/16$.
\label{fig-line16}}
\end{figure}

\begin{figure}
\plotone{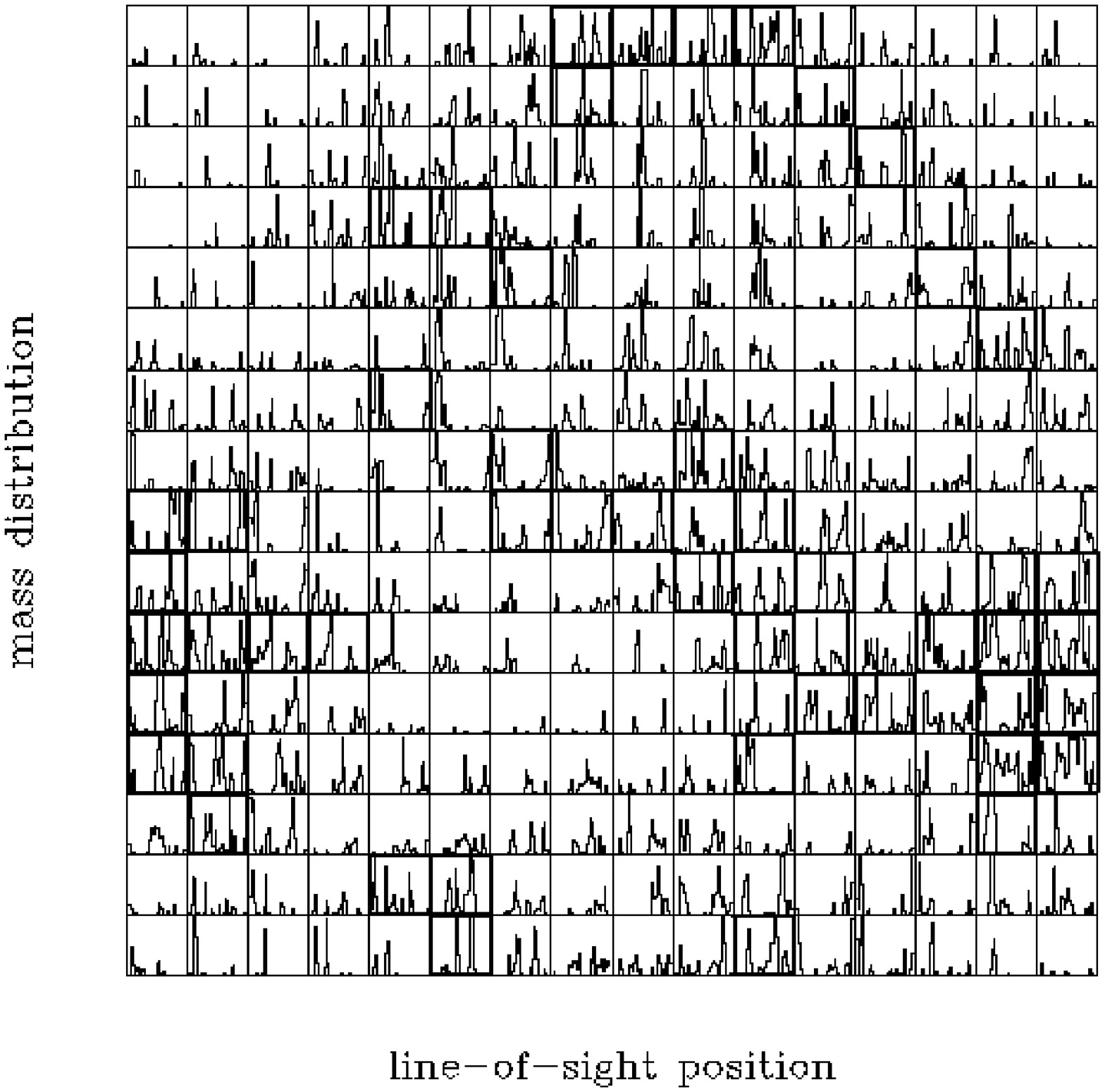}
\caption{Same as Fig. (\ref{fig-pos8}), for projected region size 
$s=L/16$.
\label{fig-pos16}}
\end{figure}

\begin{figure}
\plottwo{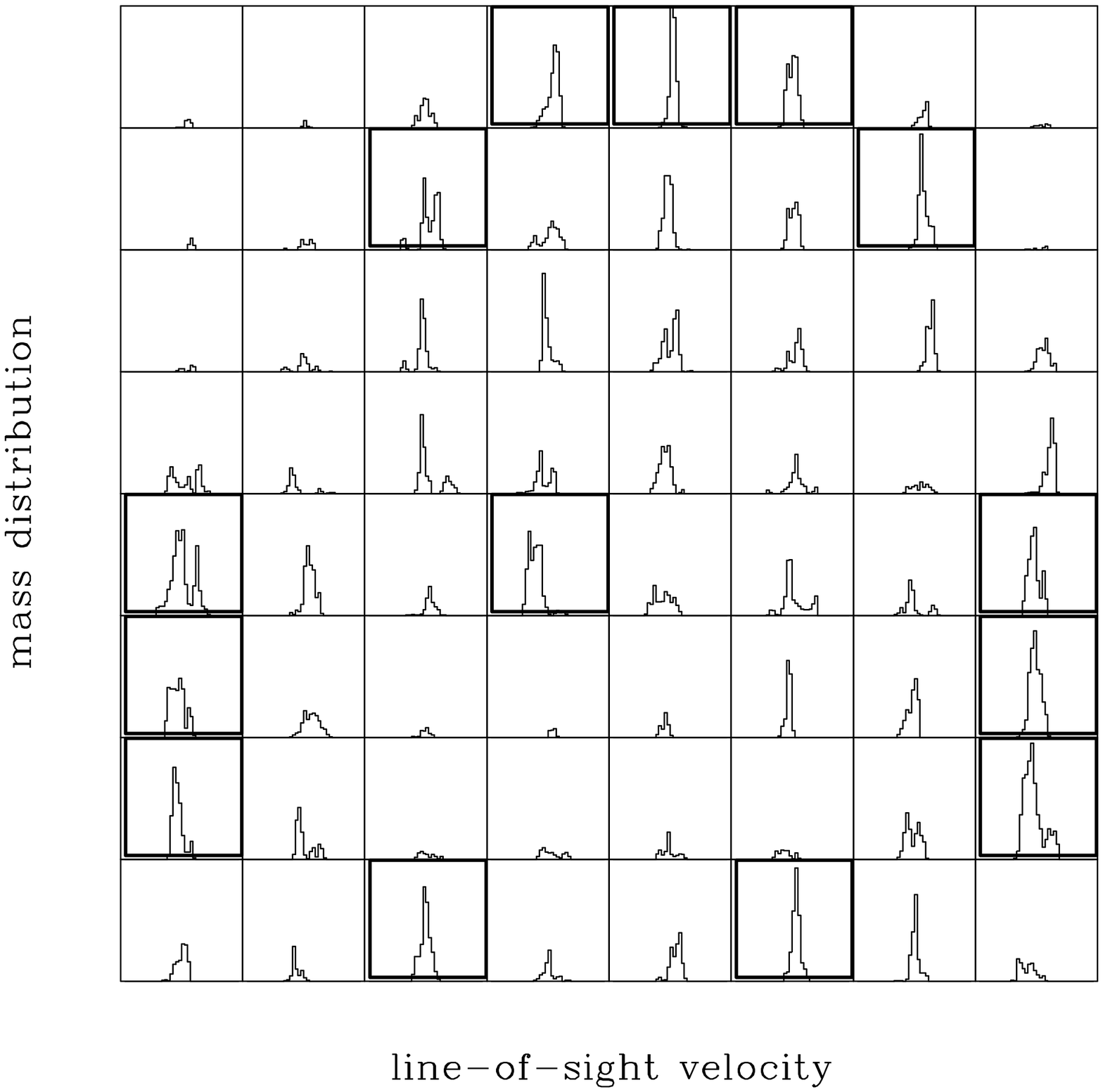}{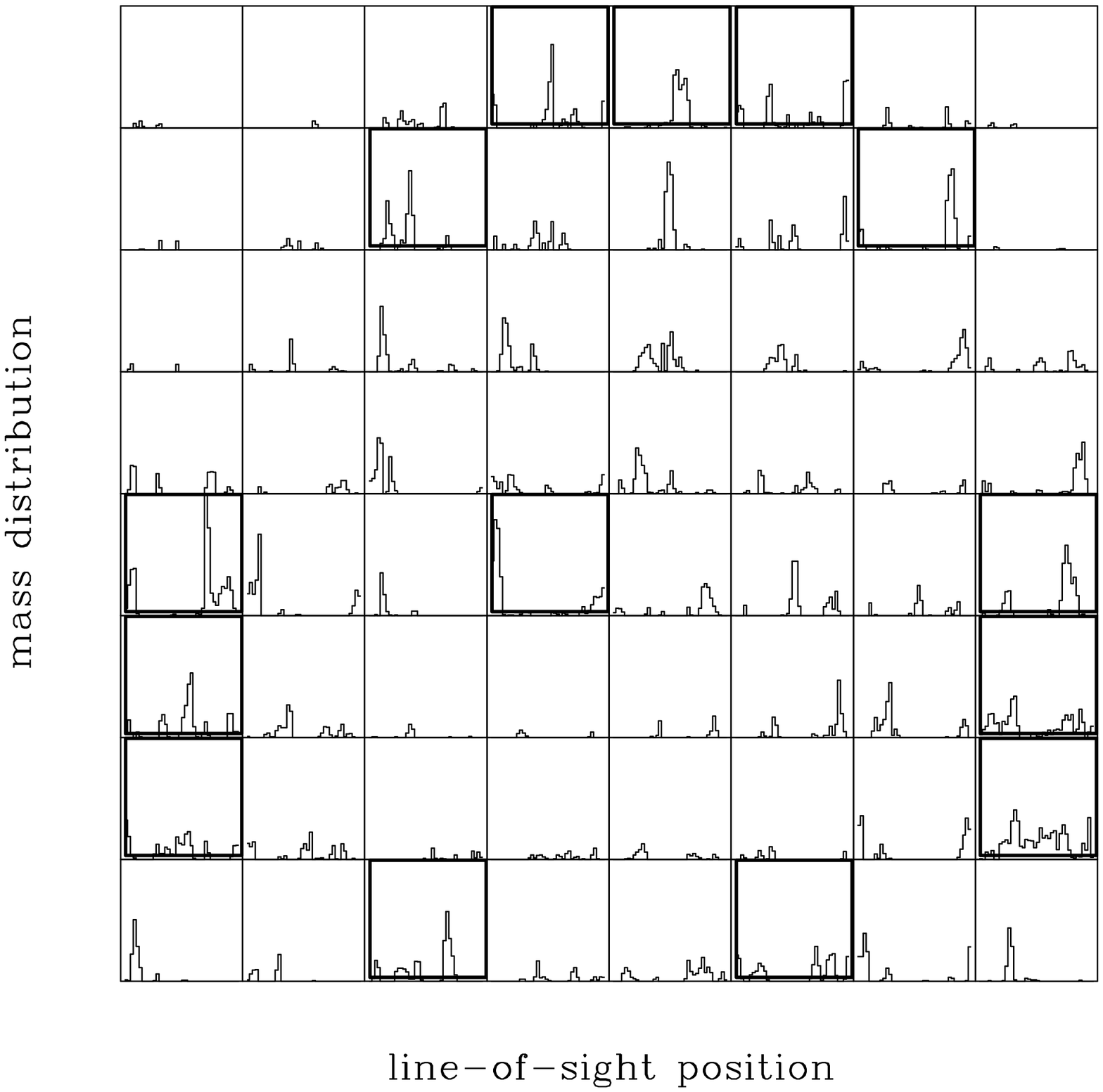}
\caption{Same as Figs. (\ref{fig-line8}), (\ref{fig-pos8}), for
$\rho_{min}=10\bar\rho$.
\label{fig-map_r10}}
\end{figure}

\begin{figure}
\epsscale{1.}
\plotone{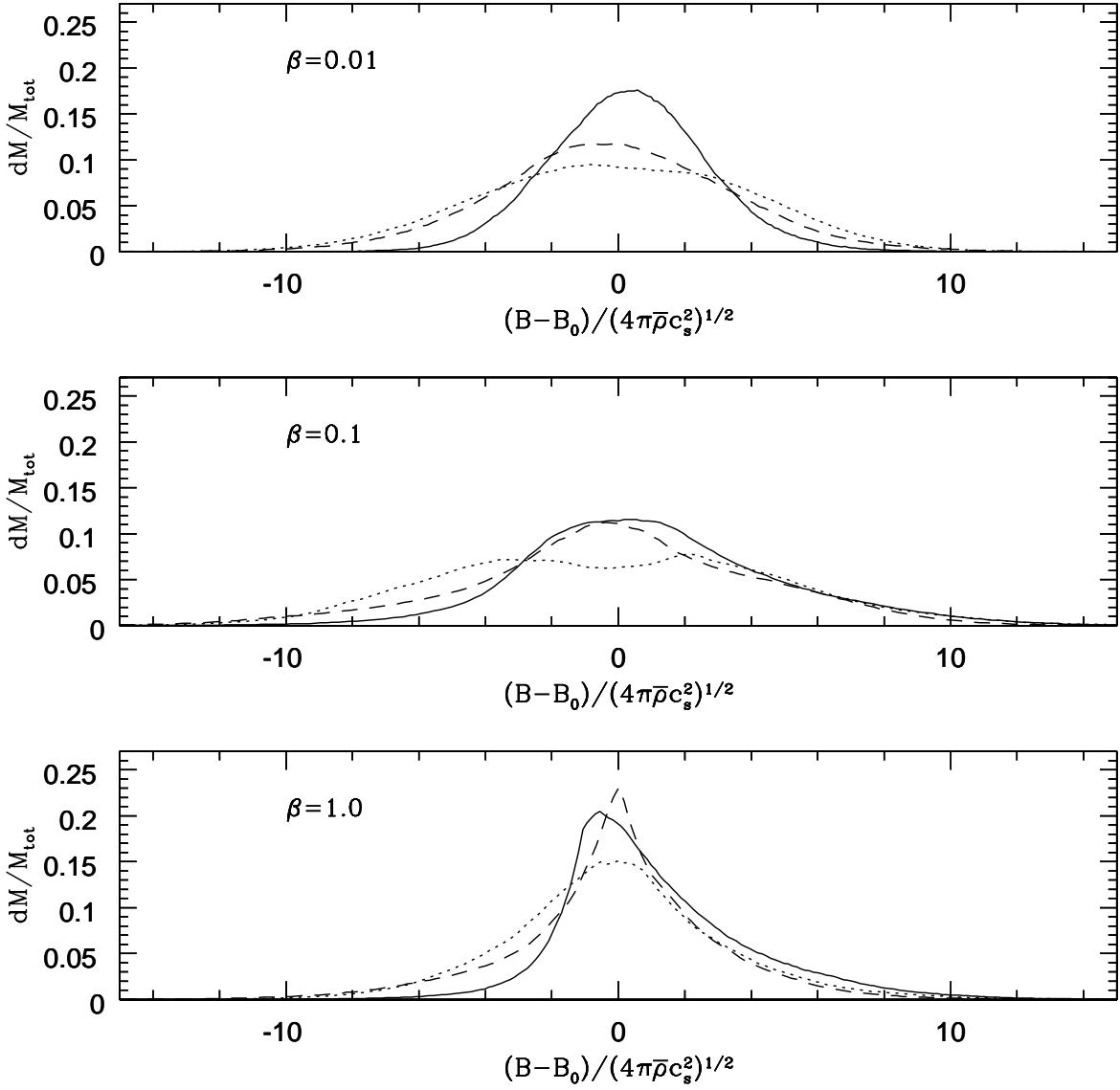}
\caption{Comparative statistics of magnetic field components for 
three model snapshots (B2, C2, D2 from Table 2) with
matched Mach numbers. Solid, dotted, and dashed curves show 
fraction of mass as a function of $B_x$, $B_y$, and $B_z$, respectively.
Mean field strengths are nonzero only in the $\hat x$ direction, with 
$B_0/(4\pi\bar\rho c_s^2)^{1/2}=10,$ 3.16, and 1, for the top, middle, and
bottom panels. 
\label{fig-bhist}}
\end{figure}

\noindent{\bf 3.}  We use a binning algorithm to investigate the
distribution of kinetic properties for plane-of-sky ``clumps'' as a
function of the spatial clump scale.  The clumps at a given scale are
simply regions that contrast with the mean surface density at that
scale; Fig. \ref{fig-roc} shows an example of the maps of these
so-defined regions of contrast (``ROCs'') at different scales.  We
find that there is a scatter in the values of line-of sight velocity
dispersion averaged over projected regions of area $s^2$, ranging from
the mean value of the velocity dispersion averaged over all cubes of
size $s^3$ up to the velocity dispersion for the entire simulation of
size $L^3$ (see Figs.  \ref{fig-lws}, \ref{fig-lws2}).  The large
velocity dispersions (``linewidths'') arise due to the superposition
of many small volume elements with differing mean velocity along lines
of sight through a given projected area.  We show that ROCs which have
single-component velocity distributions (``line profiles'') generally
consist of several spatially separated components along the
line-of-sight (see Figs.  \ref{fig-line8}-\ref{fig-map_r10}).  Thus,
what looks like a ``clump'' either on the plane of the sky or in
position-velocity datacube may in fact be a superposition of spatially
unconnected parts with smoothly overlapping velocity distributions.

\par
\begin{figure}[p]
\epsscale{1.}
\plotone{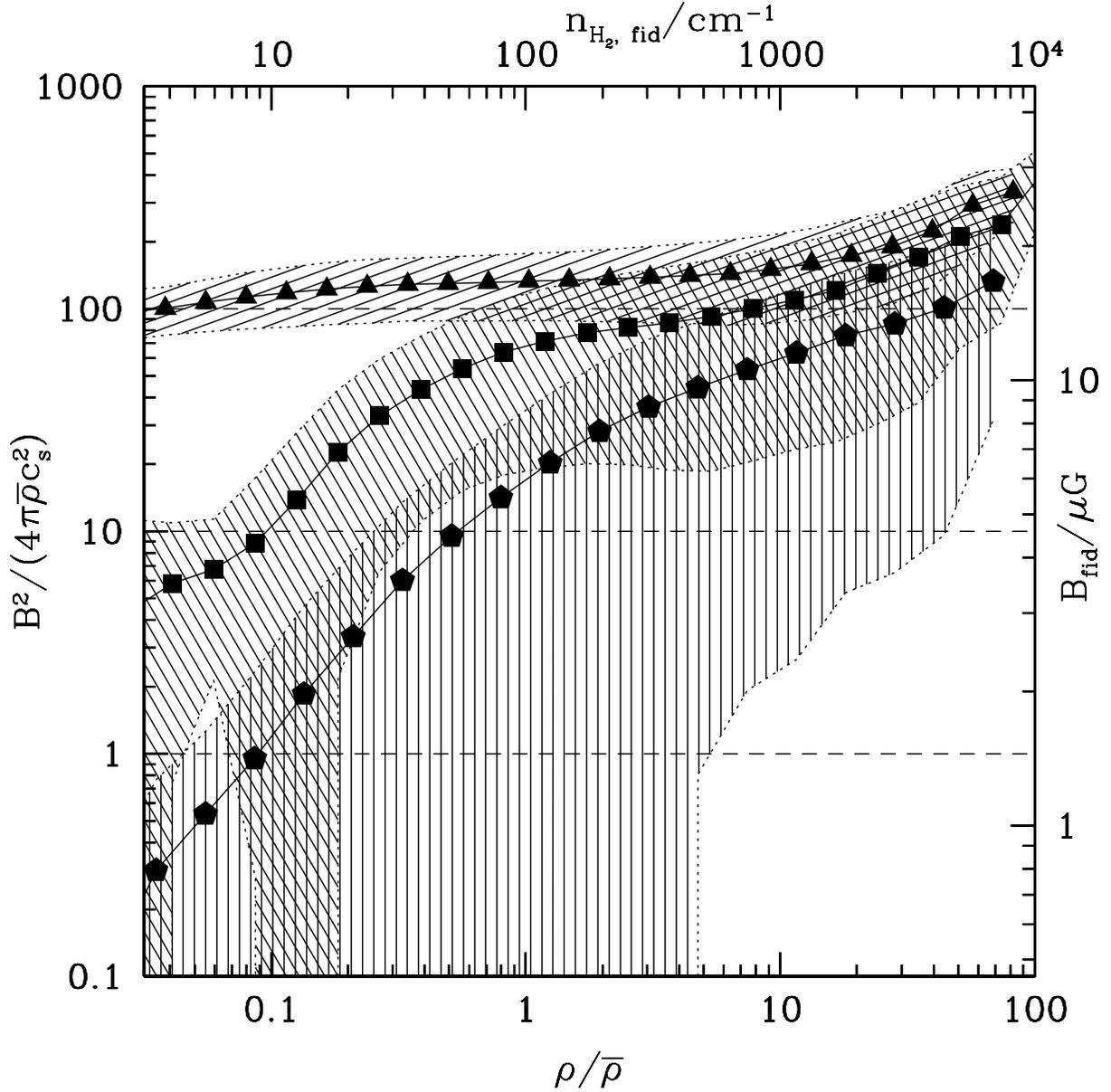}
\caption{Dependence of the total magnetic field strength on density in 
three model snapshots (B2, C2, D2 from Table 2) with
matched Mach numbers.  Triangles, squares, and pentagons show the 
mean value of $B^2$ in each density bin for Mach-7 models with 
$\beta=0.01$, $0.1$, and 1, respectively.  Shaded regions surrounding 
each curve corresponds to the $1-\sigma$ departures from the mean 
(errors in the means from counting statistics are smaller than the symbols 
shown).
Dashed horizontal 
lines show the values of the square of the mean magnetic field,
for the three models.  Left and bottom scales give 
magnetic field strength $B^2$ and density $\rho$ in dimensionless units;  
right and top scales give corresponding fiducial values of $|B|$ and 
$n_{H_2}$ assuming $T=10$K and $\bar n_{H_2}=100 \cm^{-3}$ for the 
temperature and volume-averaged density.
\label{fig-bvsrhoA}}
\end{figure}

\begin{figure}
\epsscale{1.}
\plotone{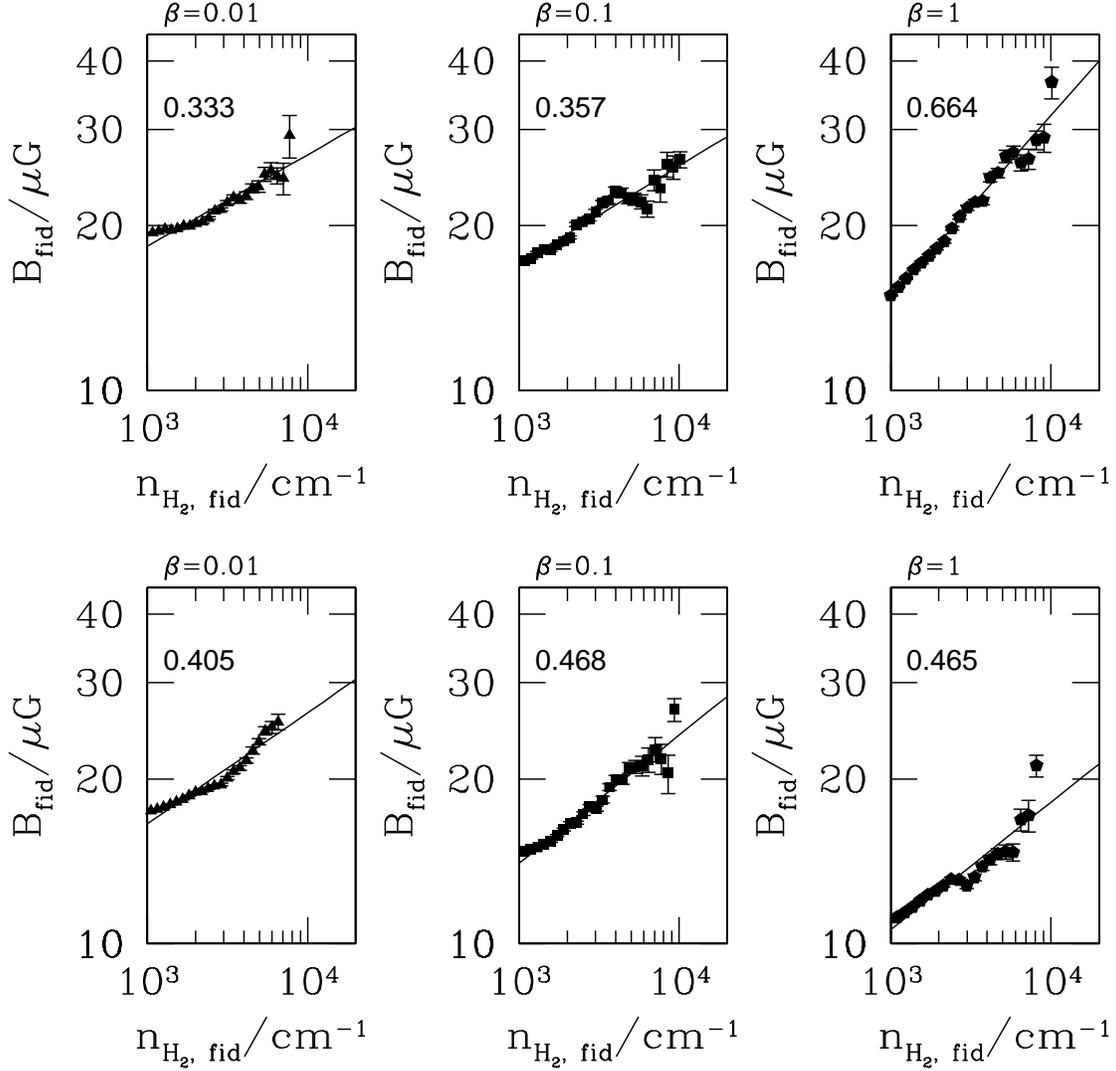}
\caption{Density vs. mean total magnetic field strength at high densities
for Mach-9 (top row) and Mach-7 (bottom row) model snapshots
(fiducial values assume $T=10$K and $\bar n_{H_2}=100 \cm^{-3}$).
Error bars show expected (Poisson noise) error in determination of the means.
Also shown (solid lines) are fits to $\log(B)$ vs. $\log(n)$ for 
$n_{H_2}>10^3 \cm^{-3}$; the corresponding slope is indicated in each panel.
\label{fig-bvrhigh}}
\end{figure}

\noindent{\bf 4.}  We find that when mean magnetic fields $B_0$ are 
weak, there is a large scatter in the distribution of total magnetic
field strength $|\bf B|$ in any given density regime, and that the
mean value of $|\bf B|$ varies strongly with density
(Fig. \ref{fig-bvsrhoA}).  For strong mean magnetic fields $B_0$ at
moderate densities, there is less scatter in the distribution of $|\bf B|$, 
and a weaker variation of the mean value of $|\bf B|$ with
density.  At high densities, the variation of the mean of $|\bf B|$
with density is similar for all the models (Fig. \ref{fig-bvrhigh}),
and is comparable to the indications of increasing field strength 
from Zeeman observations in high density tracers \citep{cru99}.

\par
\begin{figure}[p]
\epsscale{1.}
\plotone{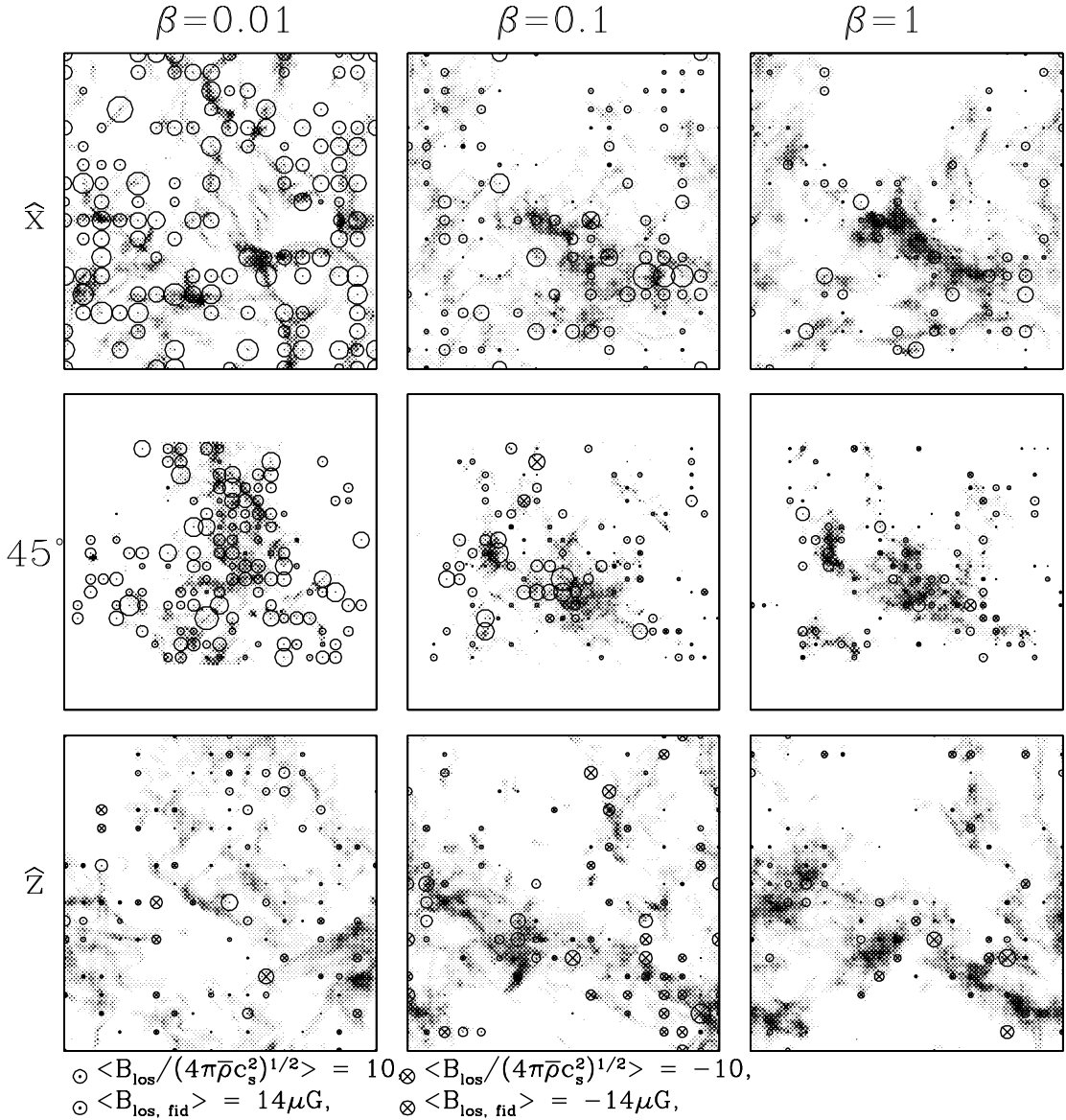}
\caption{Mean line-of-sight-averaged magnetic field strength for Mach-7 
model snapshots (B2 -- left column, C2 -- center column, D2 -- right column) 
viewed from three orientations:  along the mean field ($\hat x$; top row), at 
$45^\circ$ relative to the mean field (center row), perpendicular to the 
mean field ($\hat z$, bottom row). Contributions to $<B_{los}>$ are weighted
by density, and include only zones with $\rho/\bar{\rho}>10$.  Point size
is scaled linearly by the value of $|<B_{los}>|$, with positive and 
negative values as shown by the key.  For fiducial dimensional $B$ values, we 
adopt $T=10$K and $\bar n_{H_2}=100 \cm^{-3}$.  Only every 15th point in each
direction on the grid is plotted, for clarity.  Greyscale underlay shows the
total column density for each projection.
\label{fig-bzmap}}
\end{figure}

\begin{figure}[p]
\epsscale{1.}
\plotone{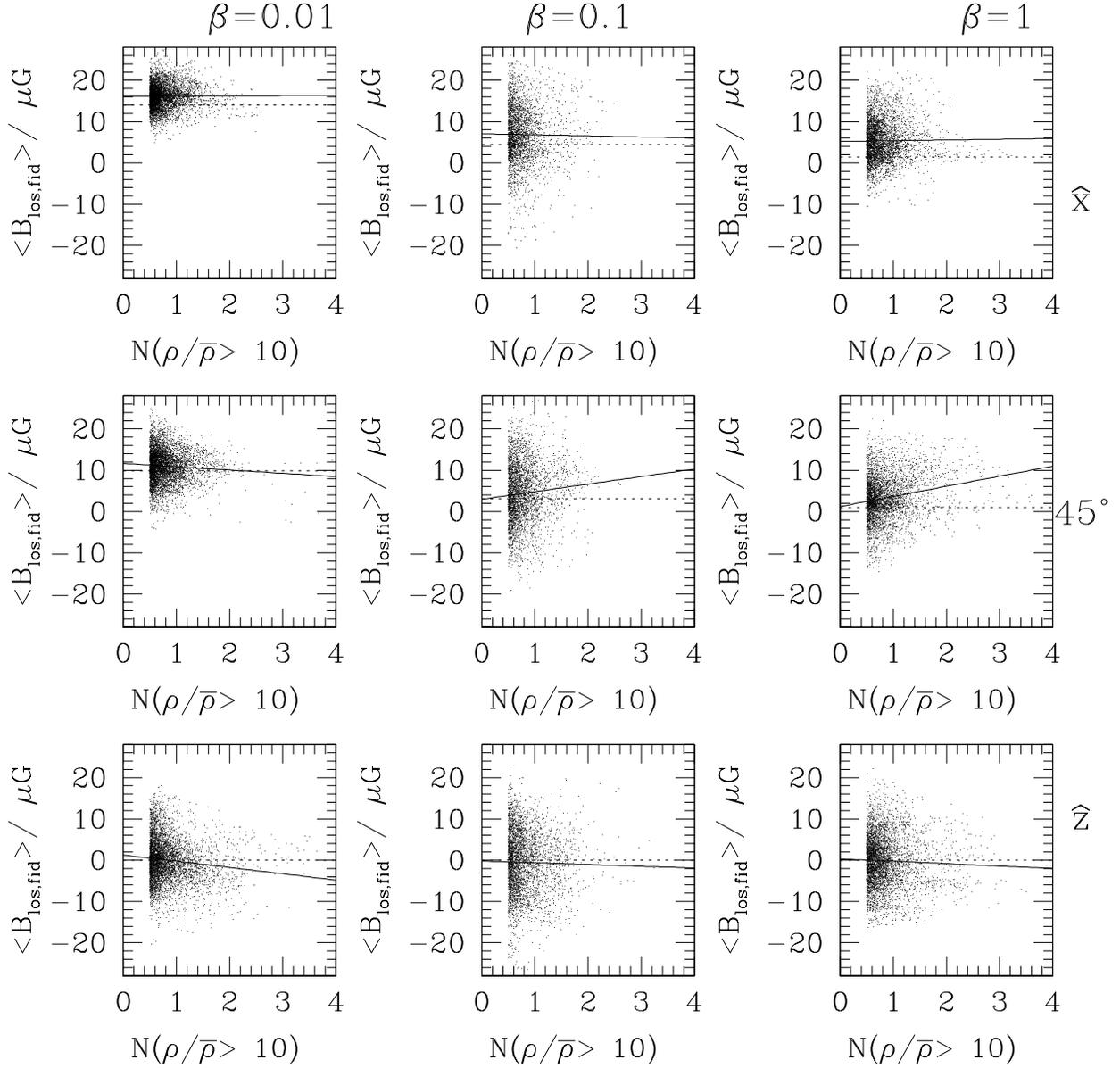}
\caption{Mean line-of-sight-averaged 
 magnetic field strength $\langle B_{los}\rangle $ vs. column 
density $N$ of gas at $\rho/\bar{\rho}>10$ for model projections shown in 
Fig. \ref{fig-bzmap}.  All points at $N/\bar{\rho}L>0.5$ are plotted;  
straight solid lines show linear least-squares fits.  Dashed horizontal 
lines show the 
value of the volume-averaged mean line-of-sight field 
$B_0\sin i$ for each projection ($i$ is the angle between the plane of 
sky and $\bf B_0$).
\label{fig-bzvscol}}
\end{figure}

\noindent{\bf 5.} We show that, for the models considered here, the
line-of-sight average of the line-of-sight magnetic field, $\langle
B_{los}\rangle$, can vary widely across a projected map, and is not
positively correlated with column density (see Figs. \ref{fig-bzmap},
\ref{fig-bzvscol}).  Because of the large scatter in $\langle
B_{los}\rangle$, an accurate observational determination of the global
average of $\langle B_{los}\rangle$ from the Zeeman method might require
a very large number of pointings.

\par
\begin{figure}
\epsscale{1.}
\plotone{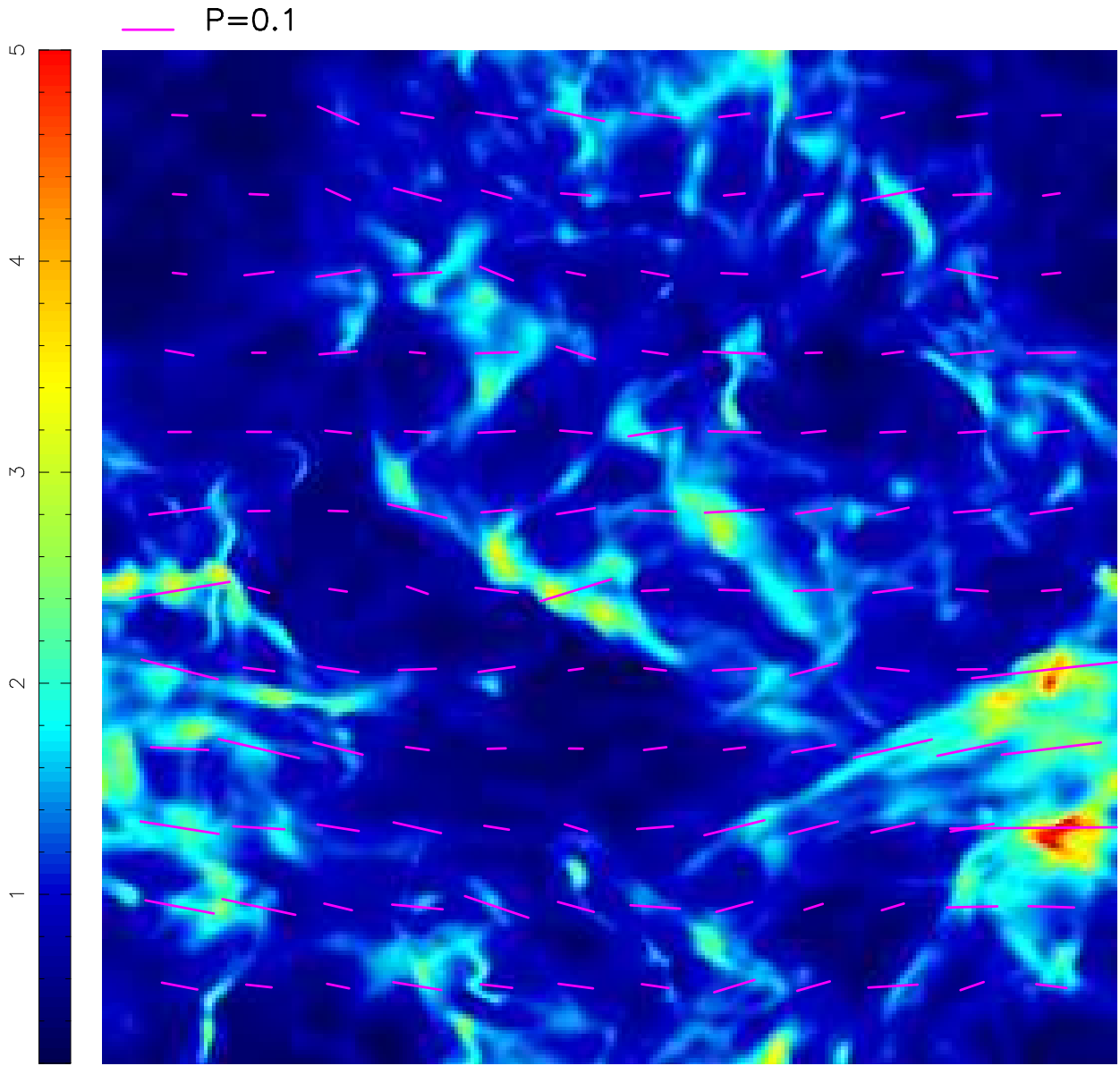}
\caption{Column density (color scale, with units $\bar{\rho}L$) 
and simulated polarization map for 
model snapshot B2 ($\beta=0.01$, ${\cal M}=7$), projected along $\hat z$ 
perpendicular to the mean magnetic field.  The fractional polarization at 
each point is proportional to the value of a fiducial polarization $P$ 
corresponding to a uniform medium and uniform magnetic 
field perpendicular to the line of sight, arbitrarily set here to $P=0.1$ as 
shown in the key.
\label{fig-imagelb}}
\end{figure}

\begin{figure}
\epsscale{1.}
\plotone{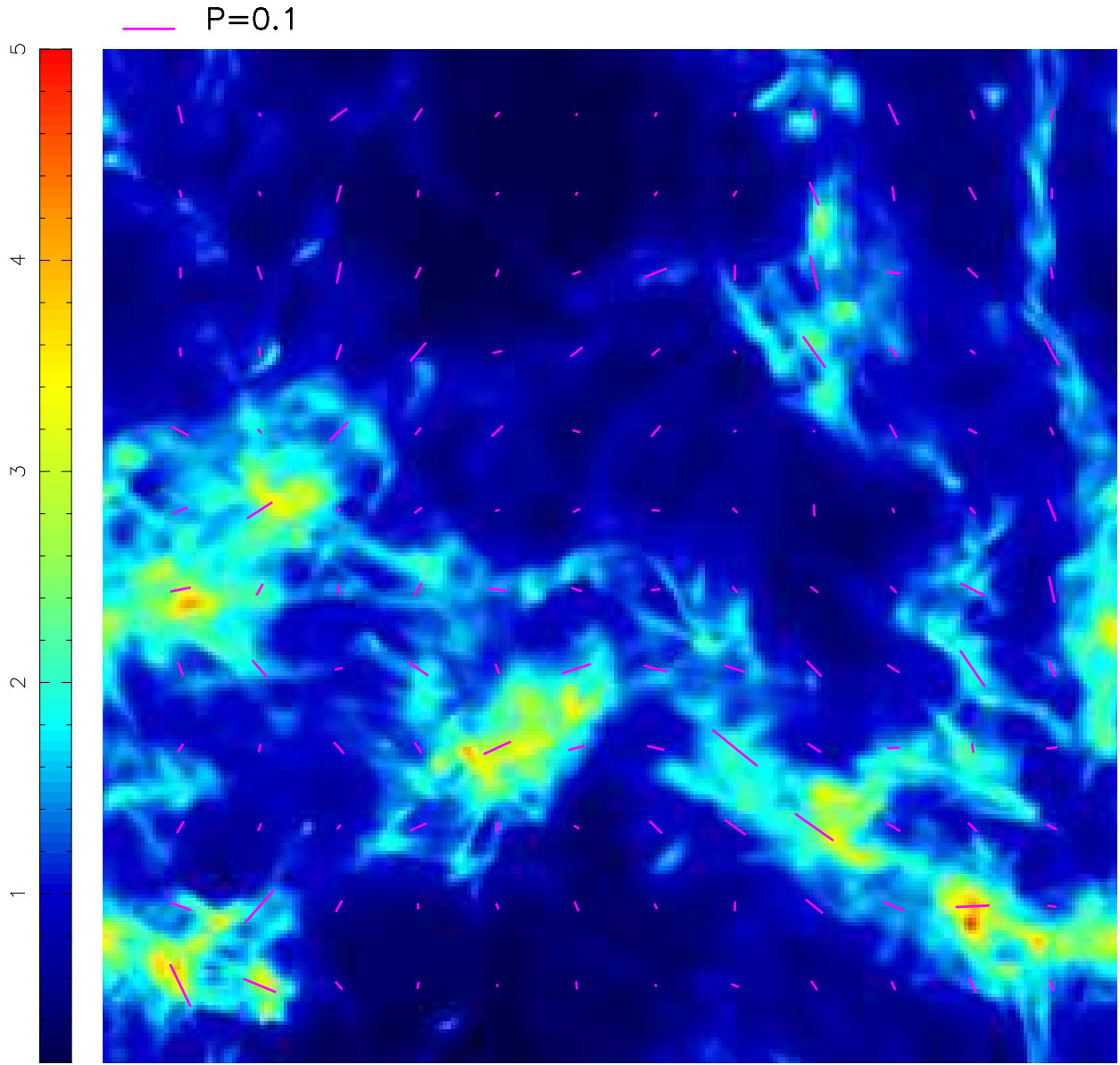}
\caption{Column density  
and polarization map for 
model snapshot D2 ($\beta=1$, ${\cal M}=7$), projected along $\hat z$;
definitions as in Fig. \ref{fig-imagelb}. 
\label{fig-imageld}}
\end{figure}

\begin{figure}
\epsscale{.9}
\plotone{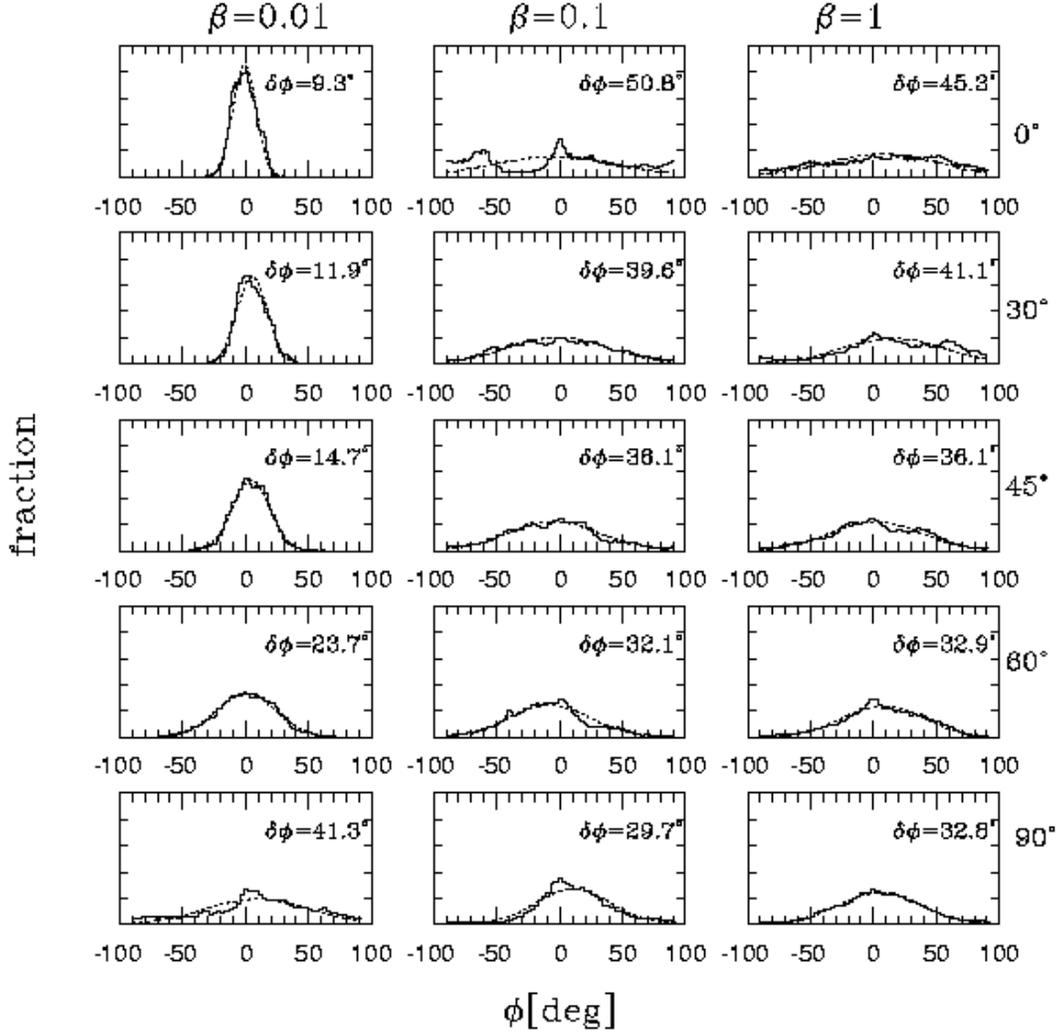}
\caption{Distribution of polarization angles for projections of models with
matched kinetic energy and differing mean magnetic field strength.  Left,
center, and right columns respectively show projections of strong, moderate,
and weak-$B_0$ cases (from ${\cal M}=7$ snapshots B2, C2, D2), for the mean 
magnetic field direction lying at varying angles with respect to the plane of 
the sky (0, 30, 45, 60, and 90$^\circ$, from top to bottom).  Each panel
shows a histogram of the distribution of polarization position angles 
with respect to the most-frequent direction, in degrees.  Labels in each panel
give the dispersion of the distribution.  Dashed curves show Gaussian fits.
\label{fig-polang}}
\end{figure}

\noindent{\bf 6.}  We have created simulated polarized-extinction
maps by integrating the Stokes parameters along lines of sight for
different simulation snapshots and orientations (Figs.\ref{fig-imagelb},
\ref{fig-imageld}).  Because models with weak $B_0$ have much more
variation in the vector direction of $\bf B$ at a given Mach number, the
result is that they have lower average values of the polarization and
larger dispersions in the polarization angle than their high-$B_0$
counterparts (Fig. \ref{fig-polang}). 
We show that the
Chandrasekhar-Fermi formula can be slightly modified to estimate
the plane-of-sky magnetic field strength $B_p$ in terms of the mean density
$\bar\rho$, 
line-of-sight velocity dispersion $\delta v_{los}$, and 
plane of sky polarization angle $\delta \phi$ as 
$B_p\approx 1.8 \bar\rho^{1/2}\delta v_{los}/\delta \phi$, 
provided $\delta \phi \simlt 25^\circ$.

One finding particularly notable for the interpretation of
molecular-line observations concerns the nature of what have long been
thought of as large-scale turbulent ``clumps'' within GMCs.  The large
scatter and relatively shallow mean slope for linewidth vs. projected
size distributions of observed ``clumps'' (e.g. \cite{wil94,stu90}) is
similar to that found for projected ``regions of contrast'' (ROCs) in
our simulated clouds.  The observed clumps are identified as coherent
structures (peaks and their surroundings) in position-velocity
molecular line data cubes; our ROCs are overdense on the plane of the
sky and generally have single-component linewidths.

Because their linewidths are comparable to those of the larger cloud
complexes in which they reside, the coherent structures in observed
$l-b-v$ cubes have been interpreted as pressure-confined clumps
(\cite{ber92}).  We believe that these apparent pressure-confined clumps
may in fact often be line-of-sight superpositions of
spatially-unconnected condensations that collectively sample from the
full turbulent velocity dispersion along the line-of-sight -- hence
appearing to have internal pressure comparable to that of the parent
cloud as a whole.  We find that the lower envelope of our linewidth-size
distribution for ROCs closely follows the relation between mean
linewidth and 3D cell size; it may be possible to apply a similar
binning procedure to observational maps in order to deduce the true 3D
mean linewidth-size relation.

Our structural analysis supports a revision (see references in \S1 and \S 5 
for related work) in the understanding of GMC
clumps that parallels the recent paradigm shift in interpreting the
cosmological Lyman-alpha forest.  Namely, we suggest that the linewidths
of these apparent clumps is due not to small-scale supersonic
``microturbulence'', but to a superposition of bulk flows with a large
range of correlation lengths.  Unlike the situation in cosmology,
however, we do not have a large-scale Hubble flow to spread out the
velocity field and help us distinguish foreground and background
concentrations.  As a consequence, what appears to be a clump in
projection may not, in three dimensions, be spatially compact or
connected at all!

A further difference with the cosmological situation is that the
overdense regions in clouds are not, in general, fated to collapse.
The overdense regions are transient objects that form, and then
disperse, from the effects of time-dependent velocity and magnetic
fields.  Eventually, through the overall dissipation of turbulence and
the random superposition of temporary concentrations, a fraction of
the material in a cloud must reach high enough densities to become
strongly self-gravitating.  The subsequent collapse and fragmentation,
perhaps initiated at many independent sites in a cloud, must
ultimately produce a collection of stars.  The next generation of
numerical simulations will require adaptive mesh algorithms to follow
this gravitational collapse and fragmentation.  Crucial questions are
whether the spectrum of stellar masses that forms can ultimately be
traced back to the bulk initial conditions (e.g. mean density, temperature,
magnetic field strength, velocity dispersion) in the parent turbulent
cloud, and what factors determine the overall rate of conversion of
gas to stars.

\acknowledgements

We gratefully acknowledge B. Elmegreen, E. Falgarone, M. Mac Low, L. Mestel,
P. Myers, \AA. Nordlund, and P. Padoan for helpful comments on the
manuscript.  We are endebted to E. V\'azquez-Semadeni for a detailed and
valuable referee's report, and to J. Alves for kindly providing a table of
extinction data for the cloud IC5146.  This work was
supported in part by NASA under grant NAG 53840, and by the National
Science Foundation under grant PHY94-07194 to the Institute for
Theoretical Physics at UCSB.  Computations were performed on the O2000
system at the National Center for Supercomputing Applications.



\clearpage


\clearpage


\clearpage

\end{document}